\documentclass[aps,prb,twocolumn]{revtex4}
\usepackage{bm}
\usepackage{tabularx}
\usepackage{ascmac}
\usepackage{amsmath}
\usepackage{wrapfig}
\usepackage{graphicx}
\usepackage{float}
\usepackage[FIGBOTCAP]{subfigure}
\usepackage{color}

\bmdefine{\bolds}{s}
\bmdefine{\boldS}{S}
\bmdefine{\boldi}{i}
\bmdefine{\boldj}{j}
\bmdefine{\boldtau}{\tau}
\bmdefine{\boldsigma}{\sigma}
\bmdefine{\boldl}{l}
\bmdefine{\boldL}{L}
\bmdefine{\boldlambda}{\lambda}
\bmdefine{\boldx}{x}
\bmdefine{\boldX}{X}
\bmdefine{\boldk}{k}
\bmdefine{\boldK}{K}
\bmdefine{\boldq}{q}
\bmdefine{\boldD}{D}
\bmdefine{\boldQ}{Q}
\bmdefine{\boldr}{r}
\bmdefine{\boldJ}{J}
\bmdefine{\boldzero}{0}
\bmdefine{\boldone}{1}
\bmdefine{\boldtwo}{2}

\begin{document}


\title{
Vector chirality for effective total momentum $J_{\textrm{eff}}$ 
in a nonfrustrated Mott insulator: 
Effects of strong spin-orbit coupling and broken inversion symmetry
}


\author{Naoya Arakawa}
\email{arakawa@hosi.phys.s.u-tokyo.ac.jp} 
\affiliation{
Center for Emergent Matter Science (CEMS), 
RIKEN, Wako, Saitama 351-0198, Japan}


\begin{abstract}
I propose the emergence of the spin-orbital-coupled vector chirality 
in a non-frustrated Mott insulator with the strong spin-orbit coupling 
due to $ab$-plane's inversion-symmetry (IS) breaking. 
I derive the superexchange interactions 
for a $t_{2g}$-orbital Hubbard model on a square lattice 
with the strong spin-orbit coupling and the IS-breaking-induced hopping integrals, 
and explain the microscopic origins of the Dzyaloshinsky-Moriya (DM) -type 
and the Kitaev-type interactions. 
Then, by adopting the mean-field approximation to 
a minimal model including only the Heisenberg-type and 
the DM-type nearest-neighbor interactions, 
I show that 
the IS breaking causes 
the spin-orbital-coupled chirality 
as a result of stabilizing the screw state. 
I also highlight the limit of the hard-pseudospin approximation 
in discussing the stability of the screw states 
in the presence of 
both the DM-type and 
the Kitaev-type interactions, and discuss its meaning. 
I finally discuss 
the effects of tetragonal crystal field and $J_{\textrm{eff}}=\frac{3}{2}$ states, 
and the application 
to the iridates near the $[001]$ surface of Sr$_{2}$IrO$_{4}$ and 
the interface between Sr$_{2}$IrO$_{4}$ and Sr$_{3}$Ir$_{2}$O$_{7}$. 

\end{abstract}


\date{\today}
\maketitle



\section{Introduction}
The spin chirality is a key concept in condensed-matter physics. 
That is categorized as either 
vector-type one, $\boldS_{\boldi}\times \boldS_{\boldj}$, 
or scalar-type one, $\boldS_{\boldi}\cdot (\boldS_{\boldj}\times \boldS_{\boldk})$. 
One of the former's drastic effects 
is to generate the electric polarization 
in multiferroic materials 
such as TbMnO$_{3}$~\cite{multiferro-exp,multiferro-Nagaosa,multiferro-theory,multiferro-1stPrinc}; 
for the latter, 
its drastic effect is to cause the anomalous-Hall effect 
in a frustrated system 
such as Pr$_{2}$Ir$_{2}$O$_{7}$~\cite{chirality-AHE-exp,chirality-AHE-Udagawa}. 
In addition to those, 
the spin chirality is related to skyrmion physics~\cite{skrymion} 
and heavy-fermion physics~\cite{chirality-HF}. 

There are two mechanisms for realizing the spin chirality. 
One arises from  
the Dzyaloshinsky-Moriya(DM)-type 
antisymmetric exchange interaction~\cite{DM1,DM2} 
such as $\textstyle\sum_{\boldi,\boldj}
\boldD_{\boldi,\boldj}\cdot (\boldS_{\boldi}\times \boldS_{\boldj})$ 
with $\boldD_{\boldi,\boldj}=-\boldD_{\boldj,\boldi}$. 
In the system in which inversion symmetry (IS) is broken 
due to lattice distortion, 
such DM-type interaction appears as a result of the combination effects of 
the onsite spin-orbit coupling (SOC), 
the kinetic exchange, and the IS breaking~\cite{DM2}; 
the DM-type interaction induces the spin vector chirality. 
This mechanism works in $\alpha$-Fe$_{2}$O$_{3}$, for example~\cite{DM2}. 
The other arises from the competition~\cite{spiral} between 
the Heisenberg-type symmetric exchange interactions 
such as $\textstyle\sum_{\boldi,\boldj}
J_{\boldi,\boldj}\boldS_{\boldi}\cdot \boldS_{\boldj}$ 
with $J_{\boldi,\boldj}=J_{\boldj,\boldi}$; 
this does not need lattice origin's IS breaking. 
Its example is MnO$_{2}$~\cite{spiral}: 
in a body-centered unit cell 
the Heisenberg-type interactions 
between the nearest-neighbor (NN) sites along the $c$ axis 
compete with others between 
the body center and each vertex; 
this competition results in stabilizing a screw state, 
where the spin vector chirality becomes finite. 

Although the understanding of the spin chirality has been developed, 
the chirality of another degree of freedom is unsatisfactorily understood. 
For example, 
we little understand the chirality of the orbital, 
although 
its possibility may be suggested by the close similarities 
between the spin and the orbital 
about orders and fluctuations~\cite{Tokura-Nagaosa}. 
Since the orbital degree of freedom of an electron describes 
the anisotropy of its spatial distribution, 
the understanding of the chirality of the orbital 
may open a new possibility of utilizing 
the chirality of the anisotropic spatial distribution of electrons/holes. 
Although there are several studies related to the chirality of the orbital, 
these focused on the frustrated iridates~\cite{hyperkagomeIr-Balents,hyperkagomeIr-Shindou,
hyperkagomeIr-Mizoguchi,honeycombIr-exp,honeycombIr-zigzag,honeycombIr-YBKim}, 
the Ir oxides with the geometric frustration of the symmetric exchange interactions. 
Since the frustration tends to develop the strong fluctuations, 
it is difficult to realize the chirality of the orbital as a result of the order; 
if it is realized, 
it may be easily broken by a small perturbation 
because such perturbation is sufficient to stabilize another competed state. 
Thus, 
the nonfrustrated system may be better 
for studies 
towards realizing the chirality related to the orbital 
and utilizing and controlling it.
However, 
it has been unclear whether the chirality related to the orbital 
is realized in the nonfrustrated system with the strong SOC 
because of lattice origin's IS breaking, 
although 
the realization may be suggested by 
the analogy with the case of the spin chirality in a nonfrustrated system 
with the weak SOC. 
Thus, 
we should clarify the possibility of the chirality related to the orbital 
in a non-frustrated system with the strong SOC in the presence of the IS breaking. 

Here I propose that 
a nonfrustrated Mott insulator with the strong SOC 
gains the spin-orbital-coupled vector chirality 
by introducing the IS breaking on an $ab$ plane 
due to the DM-type interactions 
for the spin-orbital-coupled degree of freedom~\cite{Ir214-Jeff-PRL,Ir214-Jeff-Science},  
$\boldJ_{\boldi}=\boldS_{\boldi}-\boldL_{\boldi}$ for $J_{\textrm{eff}}=1/2$. 
Focusing on the essential effects of the IS breaking 
on the low-energy physics of a quasi-two-dimensional insulating iridate, 
I derive the superexchange interactions 
for a $t_{2g}$-orbital Hubbard model 
with the strong SOC on a square lattice without $ab$-plane's IS, 
and explain why the IS breaking leads to the DM-type and 
the Kitaev-type~\cite{Kitaev} interactions. 
Then, 
by using the mean-field approximation, 
I show the emergence of the spin-orbital-coupled vector chirality 
in a minimal model with only the Heisenberg-type and the DM-type NN interactions 
as a result of stabilizing the screw state. 
Although even in the presence of the anisotropic terms such as 
the Kitaev-type interactions 
the screw state gives the lowest energy, 
the hard-pseudospin constraints are violated. 
This suggests the limit of the hard-pseudospin approximation 
in discussing the stability of the screw states 
in the presence of 
both the DM-type and the Kitaev-type interactions. 
I finally discuss 
the validity of the treatment of effects of 
tetragonal crystal field and the $J_{\textrm{eff}}=\frac{3}{2}$ states, 
and a possibility of the spin-orbital-coupled vector chirality 
near the $[001]$ surface of Sr$_{2}$IrO$_{4}$ 
and the interface between Sr$_{2}$IrO$_{4}$ and Sr$_{3}$Ir$_{2}$O$_{7}$. 
Hereafter we set $\hbar=1$ and choose the lattice constants as unity. 

\section{Method}
\subsection{Model}
\begin{figure}[tb]
\includegraphics[width=86mm]{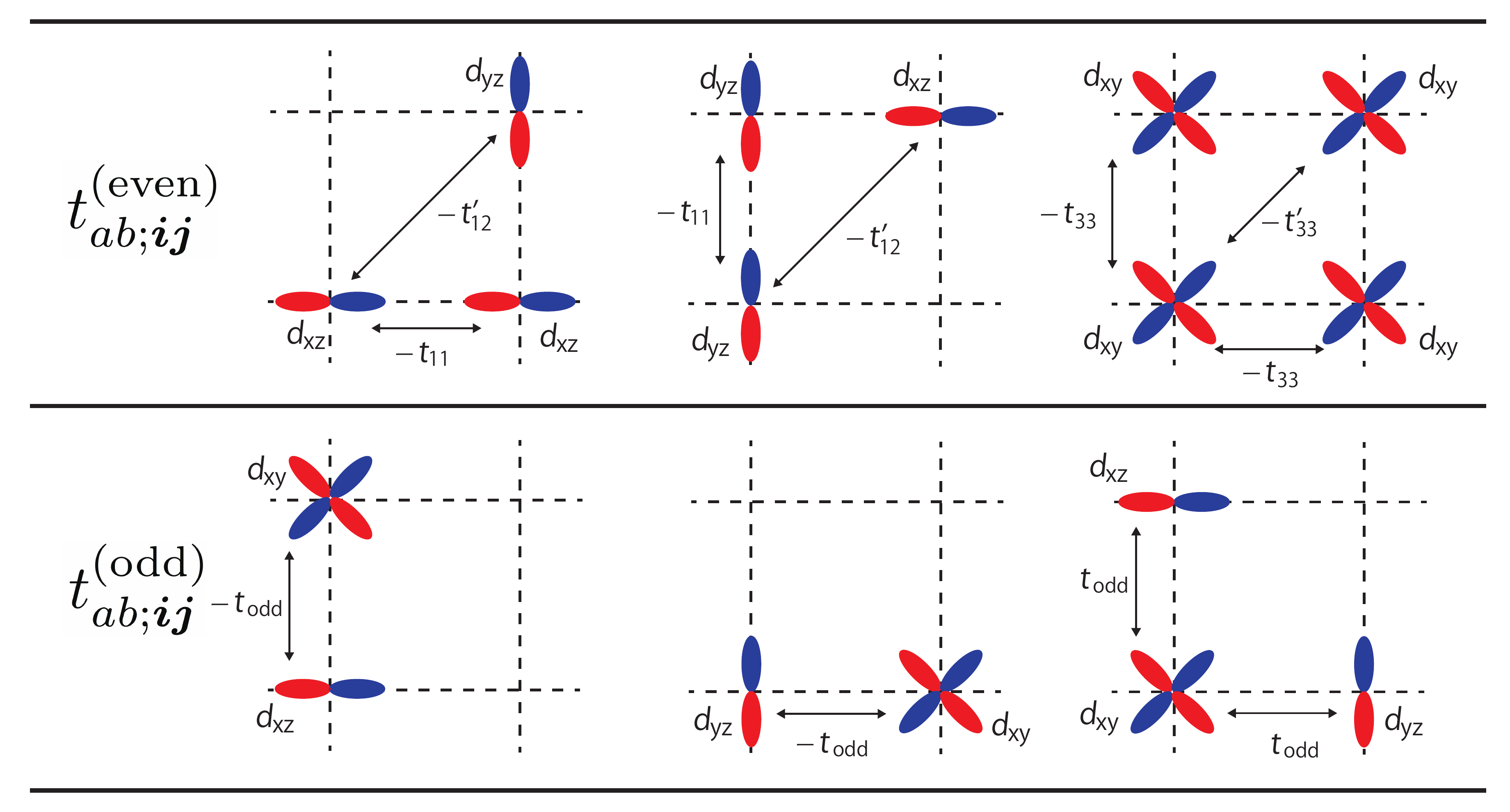}
\vspace{-6pt}
\caption{ 
Schematic pictures of finite 
$t_{ab;\boldi\boldj}^{(\textrm{even})}$ 
and $t_{ab;\boldi\boldj}^{(\textrm{odd})}$ on a square lattice. 
The NN hoppings arise from the indirect ones through the $2p$ orbitals of an O ion, 
and the next-NN hoppings arise from the direct ones. 
}
\label{fig:Fig1}
\end{figure}

We use a $t_{2g}$-orbital Hubbard model with the onsite SOC 
on a square lattice as an effective model of a quasi-two-dimensional iridate, 
and treat the effects of the IS breaking on an $ab$ plane 
as the NN hoppings~\cite{Yanase-ISB,Mizoguchi-SHE} between 
the $d_{xy}$ and $d_{yz}$ orbitals along the $x$ direction, 
and between the $d_{xy}$ and $d_{xz}$ orbitals along the $y$ direction. 
(Such treatment is appropriate for a $t_{2g}$-orbital system, 
and its applicability is wider than 
a single-orbital Rashba model~\cite{Mizoguchi-SHE}.) 
Namely, the Hamiltonian becomes
\begin{align} 
\hat{H}=\hat{H}_{\textrm{even}}+\hat{H}_{LS}+\hat{H}_{\textrm{int}}+\hat{H}_{\textrm{odd}},
\end{align} 
where $\hat{H}_{\textrm{even}}$ represents the kinetic energy, 
\begin{align}
\hat{H}_{\textrm{even}}
=\sum\limits_{\boldi,\boldj}
\sum\limits_{a,b}
\sum\limits_{s}
t_{ab;\boldi\boldj}^{(\textrm{even})}
\hat{c}_{\boldi a s}^{\dagger}\hat{c}_{\boldj b s},
\end{align}
with 
site indices, $\boldi$ and $\boldj$, for $N$ sites, 
orbital indices, $a,b=d_{xz},d_{yz},d_{xy}$, 
a spin index, $s=\uparrow,\downarrow$, 
and 
even-mirror hopping integrals, 
the hopping integrals even about $ab$-plane's mirror symmetry,  
given in Fig. \ref{fig:Fig1};  
$\hat{H}_{LS}$ the onsite SOC, 
\begin{align}
\hat{H}_{LS}
=
(-\lambda_{LS})
\sum\limits_{\boldi}
\sum\limits_{a,b}
\sum\limits_{s,s^{\prime}}
(\boldl_{\boldi}\cdot \bolds_{\boldi})_{asbs^{\prime}}
\hat{c}_{\boldi a s}^{\dagger}\hat{c}_{\boldi b s^{\prime}},
\end{align}
with the standard matrix elements $(\boldl_{\boldi}\cdot \bolds_{\boldi})_{asbs^{\prime}}$ 
(e.g., see Ref. \onlinecite{Mizoguchi-SHE}); 
$\hat{H}_{\textrm{int}}$ the multiorbital Hubbard interactions, 
\begin{align}
&\hat{H}_{\textrm{int}}
=
\sum\limits_{\boldi}\sum\limits_{a,b}
\hat{c}_{\boldi a \uparrow}^{\dagger}\hat{c}_{\boldi a \downarrow}^{\dagger}
[U\delta_{b,a}+J^{\prime}(1-\delta_{b,a})]
\hat{c}_{\boldi b \downarrow}\hat{c}_{\boldi b \uparrow}\notag\\
+&
\sum\limits_{\boldi}\sum\limits_{a,b(<a)}
\sum\limits_{s,s^{\prime}}
\hat{c}_{\boldi a s}^{\dagger}\hat{c}_{\boldi b s^{\prime}}^{\dagger}
[U^{\prime}\hat{c}_{\boldi b s^{\prime}}\hat{c}_{\boldi a s}
-J_{\textrm{H}}\hat{c}_{\boldi b s}\hat{c}_{\boldi a s^{\prime}}];
\end{align}
$\hat{H}_{\textrm{odd}}$ the Hamiltonian induced by the IS breaking~\cite{Yanase-ISB,Mizoguchi-SHE}, 
\begin{align}
\hat{H}_{\textrm{odd}}
=
\sum\limits_{\boldi,\boldj}
\sum\limits_{a,b}
\sum\limits_{s}
t_{ab;\boldi\boldj}^{(\textrm{odd})}
\hat{c}_{\boldi a s}^{\dagger}\hat{c}_{\boldj b s},
\end{align}
with 
odd-mirror hopping integral, 
the hopping integral odd about $ab$-plane's mirror symmetry, 
given in Fig. \ref{fig:Fig1}. 
In our Hamiltonian, 
we neglect the tetragonal crystal field; 
the validity will be discussed in Sec. IV. 
In the analyses, 
we consider a hole per site 
because the iridates have the $(t_{2g})^{5}$-electron configuration~\cite{Ir214-Jeff-PRL}, 
equivalent to the $(t_{2g})^{1}$-hole configuration. 

Before the derivation of the low-energy superexchange interactions, 
I briefly validate the appearance of the odd-mirror hopping integral 
in the absence of $ab$-plane's IS. 
For that purpose, 
it would be better to begin with the symmetrical properties 
of the hopping integrals permissible in a square lattice 
with $ab$-plane's IS. 
In the presence of $ab$-plane's IS, 
the permissible hopping integrals for the $t_{2g}$ orbitals 
should be even about $z$. 
Furthermore, 
the permissible hopping integrals along the $x$ and $y$ directions 
should be even about $y$ and even about $x$, respectively. 
Actually, 
those properties hold for all the hopping integrals of $\hat{H}_{\textrm{even}}$ 
because the wave functions of the $d_{xz}$, $d_{yz}$, and $d_{xy}$ orbitals 
behave like $xz$, $yz$, and $xy$, respectively, 
in symmetrical operations; 
e.g., 
the NN hopping integral of the $d_{xz}$ orbital along the $x$ direction, 
behaving like $xz\cdot xz=x^{2}y^{0}z^{2}$, 
is even about $y$ and $z$. 
Then, 
in the absence of $ab$-plane's IS, 
the hopping integrals odd about $z$, 
the odd-mirror hopping integrals, 
become permissible. 
As a result, 
the NN hopping integral between the $d_{yz}$ and the $d_{xy}$ orbitals 
along the $x$ direction is possible 
because that behaves like $yz\cdot xy=x^{1}y^{2}z^{1}$, 
which is odd about $z$ and even about $y$. 
In addition, 
the NN hopping integral between the $d_{xz}$ and the $d_{xy}$ orbitals 
along the $y$ direction is possible. 
However, 
the NN hopping integral either between the $d_{xz}$ and the $d_{xy}$ orbitals 
along the $x$ direction 
or between the $d_{yz}$ and the $d_{xy}$ orbitals along $y$ direction 
is prohibited even in the absence of $ab$-plane's IS. 
This is because the former is odd about $y$ 
and the latter is odd about $x$, 
and because $ab$-plane's IS breaking does not affect the symmetrical properties 
about $x$ and $y$. 
The above explanations are the reason why 
$ab$-plane's IS breaking induces the odd-mirror hopping integrals. 
Those hopping integrals are not only odd mirror 
but also odd parity 
because the IS breaking considered is the same at each site, 
i.e. its effects are uniform. 

\subsection{Low-energy superexchange interactions}
To understand the low-energy physics of $\hat{H}$, 
we derive the superexchange interactions~\cite{Superex-Anderson,Superex-KK} 
in a strong-correlation limit. 
Here I will show the result for $J^{\prime}=J_{\textrm{H}}=0$ and $U^{\prime}=U$ 
in $U,U^{\prime}\gg \lambda_{LS}\gg 
|t_{ab;\boldi\boldj}^{(\textrm{even})}|, 
|t_{ab;\boldi\boldj}^{(\textrm{odd})}|$~\cite{honeycombIr-zigzag,hyperkagomeIr-Mizoguchi} 
in order to focus on the essential effects of the IS breaking as simply as possible. 
Its effects on the superexchange interactions 
remain qualitatively the same as the case 
for $J^{\prime}> 0$, $J_{\textrm{H}}> 0$, and $U^{\prime} < U$ 
(see Appendix A). 
Including the effects of $\hat{H}_{LS}$ 
as the formation of the $J=\frac{1}{2}$ states~\cite{honeycombIr-zigzag}, 
rewriting $\hat{H}_{\textrm{int}}$ in terms of the irreducible representations 
of the intermediate states~\cite{Superex-titanates}, 
treating $\hat{H}_{\textrm{even}}+\hat{H}_{\textrm{odd}}$ in the second-order perturbation, 
and setting $J^{\prime}=J_{\textrm{H}}=0$ and $U^{\prime}=U$, 
we obtain an effective Hamiltonian (for more details see Appendix A): 
\begin{align}
\hspace{-3pt}
&\hat{H}_{\textrm{eff}}
=
J_{1}\hspace{-3pt}
\sum\limits_{\langle \boldi,\boldj \rangle}
\hat{\boldJ}_{\boldi}\cdot \hat{\boldJ}_{\boldj}
+J_{2}\hspace{-3pt}
\sum\limits_{\langle\langle \boldi,\boldj\rangle\rangle}
\hat{\boldJ}_{\boldi}\cdot \hat{\boldJ}_{\boldj}
+D\hspace{-3pt}
\sum\limits_{\langle \boldi,\boldj \rangle_{x}}
(\hat{\boldJ}_{\boldi}\times \hat{\boldJ}_{\boldj})^{y}\notag\\
\hspace{-3pt}
&-D\hspace{-3pt}
\sum\limits_{\langle \boldi,\boldj \rangle_{y}}
(\hat{\boldJ}_{\boldi}\times \hat{\boldJ}_{\boldj})^{x}
+K\hspace{-3pt}
\sum\limits_{\langle \boldi,\boldj\rangle_{x}}
\hat{J}_{\boldi}^{y}\hat{J}_{\boldj}^{y}
+K\hspace{-3pt}
\sum\limits_{\langle \boldi,\boldj\rangle_{y}}
\hat{J}_{\boldi}^{x}\hat{J}_{\boldj}^{x},\label{eq:Heff}
\end{align}
with 
\begin{align}
&J_{1}
=\frac{4}{9}\frac{(t_{11}+t_{33})^{2}}{U}
-\frac{16}{9}\frac{(t_{\textrm{odd}})^{2}}{U}=J_{0}-J_{\textrm{odd}},\\ 
&J_{2}=\frac{4}{9}\frac{(t_{33}^{\prime})^{2}}{U},\\ 
&D=\frac{16}{9}\frac{t_{\textrm{odd}}(t_{11}+t_{33})}{U},\\ 
&K=2J_{\textrm{odd}},
\end{align}  
$\textstyle\sum_{\langle \boldi,\boldj \rangle}
=\textstyle\sum_{\langle \boldi,\boldj \rangle_{x}}+\textstyle\sum_{\langle \boldi,\boldj \rangle_{y}}$, 
the sum of the summations taken over the NN sites along the $x$ and $y$ directions, 
and $\textstyle\sum_{\langle\langle \boldi,\boldj \rangle\rangle}$, 
the summation taken over the next-NN sites. 
Here we have neglected 
the products of the hole-density operators 
such as $\textstyle\sum_{\langle \boldi,\boldj\rangle}\hat{n}_{\boldi}\hat{n}_{\boldj}$ 
because such terms become constants in the mean-field approximation. 
Note that the finite terms of the DM-type interaction in Eq. (\ref{eq:Heff}) 
differ from the rotation-induced DM-type interaction~\cite{Ir214-weakFM-theory}. 
In Sec. IV, 
we will discuss the effect of the $J_{\textrm{eff}}=\frac{3}{2}$ states 
on the superexchange interactions. 

The derived effective Hamiltonian Eq. (\ref{eq:Heff}) 
shows three effects of the IS breaking. 
One is to cause the DM-type interactions, 
$D\textstyle\sum_{\langle \boldi,\boldj \rangle_{x}}
(\hat{\boldJ}_{\boldi}\times \hat{\boldJ}_{\boldj})^{y}$ and 
$-D\textstyle\sum_{\langle \boldi,\boldj \rangle_{y}}
(\hat{\boldJ}_{\boldi}\times \hat{\boldJ}_{\boldj})^{x}$. 
Those interactions arise from 
the multiorbital superexchange interactions due to the combination 
of the even-mirror and the odd-mirror hopping integrals. 
Such combination 
is vital to obtain the DM-type antisymmetric interactions. 
This is because their operator parts 
behave like the functions odd about some coordinates 
in the symmetrical operations. 
For example, 
$(\hat{\boldJ}_{\boldi}\times \hat{\boldJ}_{\boldj})^{x}
=\hat{J}^{y}_{\boldi}\hat{J}^{z}_{\boldj}-\hat{J}^{z}_{\boldi}\hat{J}^{y}_{\boldj}$ 
behaves like the function which 
is odd about $y$ and $z$ (and even about $x$); 
such function can be obtained by the multiorbital superexchange interactions 
using 
the even-mirror hopping integral and 
the odd-mirror hopping integral 
between the $d_{xz}$ and the $d_{xy}$ orbitals, 
which behaves like $xz\cdot xy =x^{2}y^{1}z^{1}$. 
Thus, 
the DM-type interactions originate from the mirror-mixing multiorbital effect. 
Another effect is to cause the ferromagnetic Heisenberg-type interaction, 
$-J_{\textrm{odd}}\textstyle\sum_{\langle \boldi,\boldj\rangle}\hat{\boldJ}_{\boldi}\cdot\hat{\boldJ}_{\boldj}$. 
This interaction between the $z$ components 
arises from a larger gain of the energy reduction 
due to the kinetic exchange 
between the same-$J^{z}$ states 
than between the opposite-$J^{z}$ states. 
This is because the number of processes in the former case 
is larger due to the opposite spin indices between 
the $d_{xy}$ orbital and the $d_{xz}$ or $d_{yz}$ orbital 
in the $J^{z}=\pm \frac{1}{2}$ states~\cite{Rigand-text}. 
Then, 
by including the pseudospin-flipping processes, 
we can obtain the ferromagnetic Heisenberg-type interaction 
because these processes give the $\hat{J}_{\boldi}^{+}\hat{J}_{\boldj}^{+}$ 
or $\hat{J}_{\boldi}^{-}\hat{J}_{\boldj}^{-}$ terms. 
The other effect is to cause the antiferromagentic Kitaev-type interactions, 
$K\textstyle\sum_{\langle \boldi,\boldj\rangle_{x}}\hat{J}_{\boldi}^{y}\hat{J}_{\boldj}^{y}$ 
and $K\textstyle\sum_{\langle \boldi,\boldj\rangle_{y}}\hat{J}_{\boldi}^{x}\hat{J}_{\boldj}^{x}$. 
This is because the above $J_{\boldi}^{+}J_{\boldj}^{+}$ 
or $\hat{J}_{\boldi}^{-}\hat{J}_{\boldj}^{-}$ terms give the extra $\hat{J}_{\boldi}^{x}\hat{J}_{\boldj}^{x}$ 
or $J_{\boldi}^{y}J_{\boldj}^{y}$ terms, 
and because the orbitals hybridized by $\hat{H}_{\textrm{odd}}$ 
along the $x$ and $y$ direction have 
the finite off-diagonal matrix elements 
only of $\hat{L}_{\boldi}^{y}$ or $\hat{L}_{\boldi}^{x}$, respectively. 
Namely, 
the superexchange interactions 
between $\hat{L}_{\boldi}^{y}$ and $\hat{L}_{\boldj}^{y}$ along the $x$ direction 
and between $\hat{L}_{\boldi}^{x}$ and $\hat{L}_{\boldj}^{x}$ along the $y$ direction 
become antiferromagnetic 
in order to gain the energy reduction 
of the kinetic exchange due to the spin-independent interorbital hoppings 
of $\hat{H}_{\textrm{odd}}$, 
resulting in the antiferromagnetic Kitaev-type interactions. 

\subsection{Mean-field approximation}
For further understanding of the effects of the IS breaking, 
we analyze the ground state of Eq. (\ref{eq:Heff}) in the mean-field approximation. 
Since the energy in this approximation 
is quadratic about 
$\langle \hat{J}_{\boldq}^{\alpha}\rangle
=\frac{1}{\sqrt{N}}\textstyle\sum_{\boldj}e^{-i\boldq\cdot \boldj}
\langle \hat{J}_{\boldj}^{\alpha}\rangle$, i.e.
\begin{align}
\langle \hat{H}_{\textrm{eff}}\rangle 
=\sum\limits_{\boldq}
\sum\limits_{\alpha,\beta=x,y,z}
J_{\alpha\beta}(\boldq)
\langle \hat{J}_{\boldq}^{\alpha}\rangle^{\ast}
\langle \hat{J}_{\boldq}^{\beta}\rangle,\label{eq:Heff-MFA}
\end{align}
we can determine the ground state by 
finding $\boldq$ of the lowest eigenvalue and the eigenvector of Eq. (\ref{eq:Heff-MFA}) 
with the periodic boundary condition 
and the constraints of the hard-pseudospin approximation,
\begin{align}
\frac{1}{4}
=&\frac{1}{N}\sum\limits_{\alpha}
\sum\limits_{\boldq,\boldq^{\prime}}
e^{i(\boldq^{\prime}-\boldq)\cdot \boldj}
\langle \hat{J}_{\boldq}^{\alpha}\rangle^{\ast}
\langle \hat{J}_{\boldq^{\prime}}^{\alpha}\rangle\label{eq:constraint-q},
\end{align}
in which $\langle \hat{J}_{\boldi}^{\alpha}\rangle$ for all $N$ sites 
are treated as the hard pseudospins for $J_{\textrm{eff}}=\frac{1}{2}$. 
In Eq. (\ref{eq:Heff-MFA}), 
$J_{\alpha\beta}(\boldq)$ are given by 
\begin{align}
J_{xx}(\boldq)=&
(J_{0}-J_{\textrm{odd}})\cos q_{x}
+(J_{0}+J_{\textrm{odd}})\cos q_{y}\notag\\
&+4J_{2}\cos q_{x}\cos q_{y},\label{eq:Jqxx}\\
J_{yy}(\boldq)=&
(J_{0}+J_{\textrm{odd}})\cos q_{x}
+(J_{0}-J_{\textrm{odd}})\cos q_{y}\notag\\
&+4J_{2}\cos q_{x}\cos q_{y},\\
J_{zz}(\boldq)=&
(J_{0}-J_{\textrm{odd}})\cos q_{x}
+(J_{0}-J_{\textrm{odd}})\cos q_{y}\notag\\
&+4J_{2}\cos q_{x}\cos q_{y},\\
J_{zx}(\boldq)=&\ i D\sin q_{x},\\
J_{xz}(\boldq)=&-i D\sin q_{x},\\
J_{zy}(\boldq)=&\ i D \sin q_{y},\\
J_{yz}(\boldq)=&-i D \sin q_{y}\label{eq:Jqyz}.
\end{align}
The derivation of Eq. (\ref{eq:Heff-MFA}) 
from Eq. (\ref{eq:Heff}) is described in Appendix B. 

\section{Results}
Before the analyses with the IS breaking, 
we briefly explain the ground state 
without the IS breaking. 
Without it, 
the ground state is determined by the $J_{0}$ term and the $J_{2}$ term 
of Eq. (\ref{eq:Heff}).
Namely, 
the ground state is a $(\pi,\pi)$-antiferromagnetic state for $J_{0}> 4J_{2}$, 
and a $(\pi,0)$- or $(0,\pi)$-antiferromagnetic state for $J_{0}< 4J_{2}$.
The former is realized in a realistic case 
because $J_{0}> 4J_{2}$ is satisfied 
due to $|t_{11}|\sim |t_{33}|$ and $|t_{33}^{\prime}|< |t_{33}|$~\cite{Ir214-Watanabe}. 

\subsection{Stability of the screw states}
To understand how the IS breaking affects 
the $(\pi,\pi)$-antiferromagnetic state, 
realized without the IS breaking, 
let us consider a minimal model 
with only the $J_{0}$ term and the $D$ terms of Eq. (\ref{eq:Heff}). 
This minimal model is reasonable to analyze the essential effects 
of the IS breaking 
because $D$ is larger than $K$ for $2|t_{\textrm{odd}}| < |t_{11}+t_{33}|$, 
corresponding to the case with the small effects of the IS breaking. 
(Such small-effect case is considered 
because the small effects are realized as long as the IS is broken.) 
For this minimal model, 
the lowest eigenvalue of Eq. (\ref{eq:Heff-MFA}) for each $\boldq$
is determined by
\begin{align} 
\lambda(\boldq)=J_{0}(\cos q_{x}+\cos q_{y})-|D|\sqrt{(\sin q_{x})^{2}+(\sin q_{y})^{2}},
\end{align}
and it has three kinds of local minimum: 
(i) the $(\pi,\pi)$-antiferromagnetic state,
\begin{align}
\lambda(\boldQ_{\textrm{AF}})=-2J_{0}
\end{align} 
with $\boldQ_{\textrm{AF}}=(\pi,\pi)$;
(ii) the one-directional-screw state,
\begin{align} 
\lambda(\boldQ_{1})=-J_{0}-\sqrt{J_{0}^{2}+D^{2}} 
\end{align}
with $\boldQ_{1}=(\pm Q_{1},\pi)$ or $\boldQ_{1}=(\pi,\pm Q_{1})$ 
for $Q_{1}=\pi-\cos^{-1}(J_{0}/\sqrt{J_{0}^{2}+D^{2}})$;
and (iii) the two-directional-screw state,
\begin{align} 
\lambda(\boldQ_{2})=-\sqrt{2}\sqrt{2J_{0}^{2}+D^{2}} 
\end{align}
with $\boldQ_{2}=(\pm Q_{2},\pm Q_{2})$ or $\boldQ_{2}=(\pm Q_{2},\mp Q_{2})$
for $Q_{2}=\pi-\cos^{-1}(\sqrt{2}J_{0}/\sqrt{2J_{0}^{2}+D^{2}})$.

\begin{figure}[tb]
\includegraphics[width=60mm]{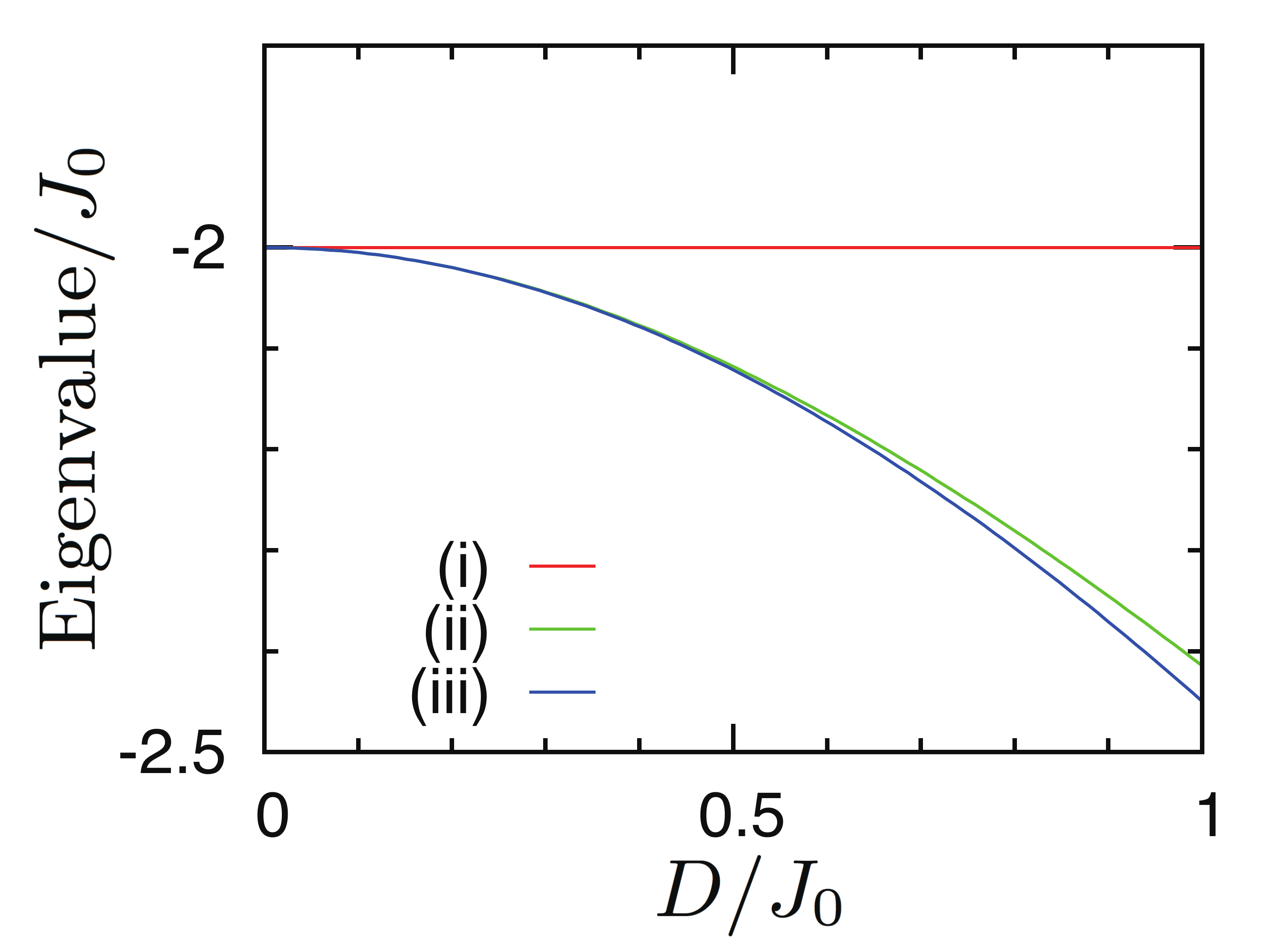}
\vspace{-6pt}
\caption{ 
$(D/J_{0})$ dependences of the eigenvalues divided by $J_{0}$ 
for states (i), (ii), and (iii) 
in the mean-field approximation for the minimal model. 
}
\label{fig:Fig2}
\end{figure}
In contrast to the $(\pi,\pi)$-antiferromagnetic state, 
those screw states 
show two unusual features. 
One is the spatial-dependent mixing between or among 
the components of $\langle \hat{J}_{\boldi}^{\alpha}\rangle$: 
for example, 
in the one-directional-screw state for $\boldQ_{1}=(\pm Q_{1},\pi)$, 
\begin{align}
\langle \hat{\boldJ}_{\boldi}\rangle
=\frac{1}{2}
\left(
\begin{array}{@{\,}c@{\,}}
(-1)^{i_{y}+1}\textrm{sgn}(D)\sin Q_{1}i_{x}\\
0\\
(-1)^{i_{y}}\cos Q_{1}i_{x}
\end{array}
\right);
\end{align} 
and in the two-directional-screw state for $\boldQ_{2}=(\pm Q_{2},\pm Q_{2})$, 
\begin{align}
\langle \hat{\boldJ}_{\boldi}\rangle
=\frac{1}{2}
\left(
\begin{array}{@{\,}c@{\,}}
-\frac{1}{\sqrt{2}}\textrm{sgn}(D)\sin \boldQ_{2}\cdot \boldi\\
-\frac{1}{\sqrt{2}}\textrm{sgn}(D)\sin \boldQ_{2}\cdot \boldi\\
\cos \boldQ_{2}\cdot \boldi
\end{array}
\right).
\end{align} 
Thus, 
those screw states have the spatial variation of not only the spin distribution 
but also the orbital distribution 
because  
the spin and the orbital are highly entangled 
in the $J_{\textrm{eff}}=\frac{1}{2}$ states~\cite{Rigand-text} 
[see Eqs. (\ref{eq:Jzp}) and (\ref{eq:Jzm})]. 
The other is the finite vector chirality: 
for example, 
in the one-directional-screw state for $\boldQ_{1}=(\pm Q_{1},\pi)$, 
the finite term is 
\begin{align}
\langle (\hat{\boldJ}_{\boldi}\times \hat{\boldJ}_{\boldj})^{y}\rangle
=\frac{1}{4}(-1)^{i_{y}+j_{y}}\textrm{sgn}(D)\sin Q_{1}(j_{x}-i_{x});
\end{align} 
and in the two-directional-screw state for $\boldQ_{2}=(\pm Q_{2},\pm Q_{2})$, 
the finite terms are 
\begin{align} 
\langle (\hat{\boldJ}_{\boldi}\times \hat{\boldJ}_{\boldj})^{x}\rangle
=\frac{1}{4\sqrt{2}}\textrm{sgn}(D)\sin \boldQ_{2}\cdot (\boldj-\boldi)
\end{align} 
and 
\begin{align}
\langle (\hat{\boldJ}_{\boldi}\times \hat{\boldJ}_{\boldj})^{y}\rangle
=-\frac{1}{4\sqrt{2}}\textrm{sgn}(D)\sin \boldQ_{2}\cdot (\boldj-\boldi).
\end{align} 
Thus, 
those screw states have the spin-orbital-coupled vector chirality. 

From the $(D/J_{0})$ dependences of the eigenvalues of states (i), (ii), and (iii), 
shown in Fig. \ref{fig:Fig2}, 
we find that 
the screw states are more stable than the $(\pi,\pi)$-antiferromagnetic state, 
and that 
the most stable state in the minimal model 
is the two-directional-screw state. 
The stabilities of those screw states can be understood that 
the $J_{0}$ term becomes minimum for $\boldq=(\pi,\pi)$, 
and the $D$ terms become minimum for $\boldq=(\frac{\pi}{2},\frac{\pi}{2})$. 

\subsection{Limit of the hard-pseudospin approximation}
\begin{figure}[tb]
\includegraphics[width=60mm]{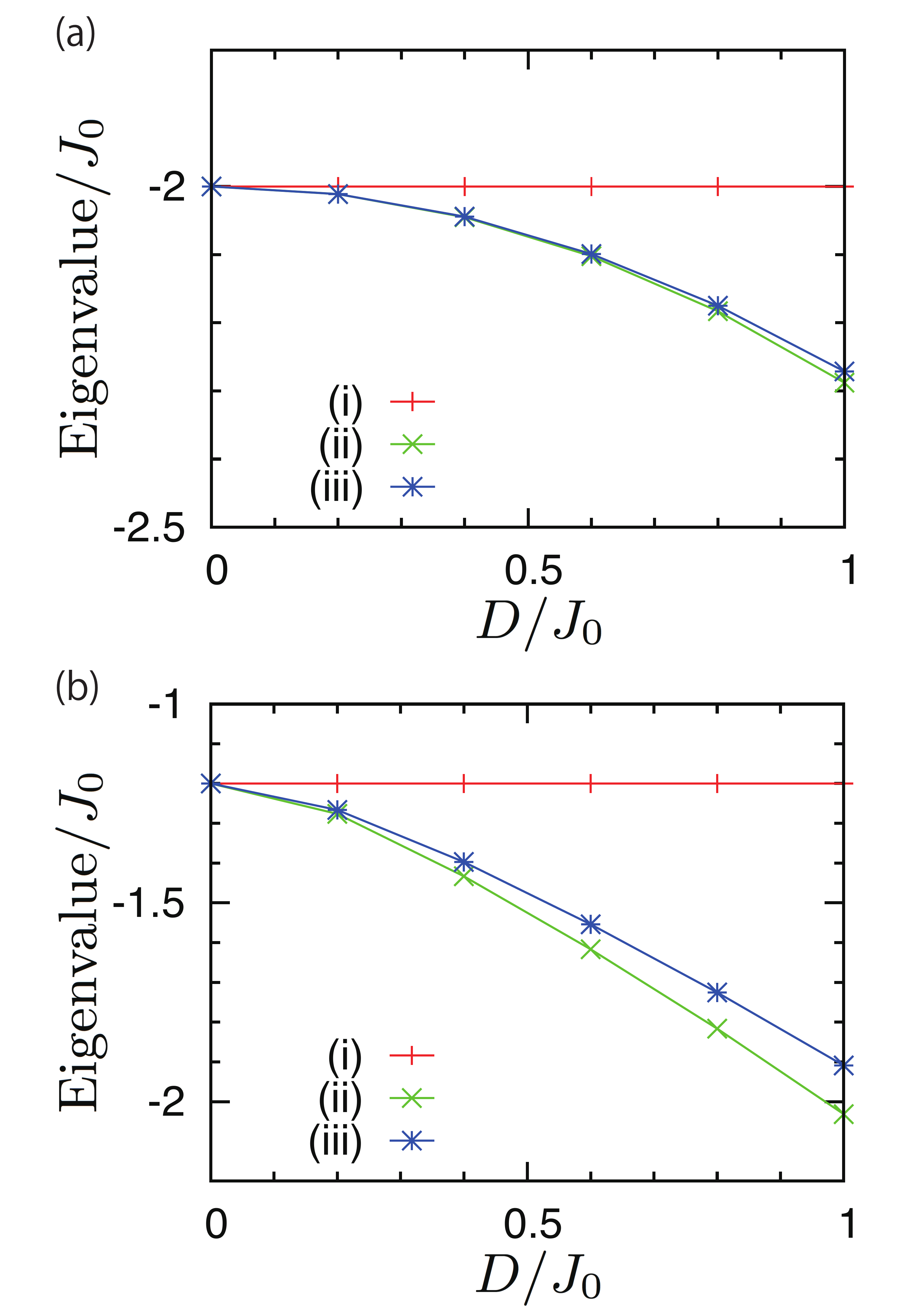}
\vspace{-6pt}
\caption{
$(D/J_{0})$ dependences of the eigenvalues divided by $J_{0}$ 
for states (i), (ii), and (iii) 
in the mean-field approximation with $N=100\times 100$ meshes 
for the model with the $J_{0}$ and the $D$ terms, 
$J_{\textrm{odd}}=\frac{1}{4}J_{0}(\frac{D}{J_{0}})^{2}$, and 
$J_{2}=$ (a) $0$ or (b) $0.2J_{0}$. 
}
\label{fig:Fig3}
\end{figure}
Even in the presence of the $J_{2}$ term and the $J_{\textrm{odd}}$ terms, 
the eigenvalues of the screw states are lower than 
the eigenvalue for $\boldq=(\pi,\pi)$, 
as shown in Figs. \ref{fig:Fig3}(a) and \ref{fig:Fig3}(b). 
In particular, 
the one-directional-screw state 
for $\boldQ_{1}^{\prime}=(\pm Q_{1}^{\prime},\pi)$ or $\boldq=(\pi,\pm Q_{1}^{\prime})$ 
gives the lowest eigenvalue (see those figures). 
This is because the $J_{\textrm{odd}}$ terms and the $J_{2}$ term destabilize 
the states for $\boldq=(Q,\pm Q)$. 
Namely, 
the lowest eigenvalue of the one-directional-screw state 
arises from the combination 
of the stabilization of the screw states due to the $D$ terms 
and the destabilization of the two-directional-screw state 
(compared with the one-directional-screw state) 
due to the $J_{\textrm{odd}}$ terms and the $J_{2}$ term. 

In contrast to the case of the minimal model, 
the screw states in the presence of the $J_{\textrm{odd}}$ terms 
do not satisfy the hard-pseudospin constraints of $\langle \hat{\boldJ}_{\boldj}\rangle$. 
This is because under the constraints, 
the possible one-directional-screw state is restricted to 
$\langle \hat{\boldJ}_{\boldQ}\rangle={}^{t}(\pm ib, 0, b)$ 
or $\langle \hat{\boldJ}_{\boldQ}\rangle={}^{t}(0,\pm ib, b)$, 
although 
in the presence of the anisotropic terms such as the $J_{\textrm{odd}}$ terms
the realized state becomes 
$\langle \hat{\boldJ}_{\boldQ}\rangle={}^{t}(\pm ib, 0, c)$ 
or $\langle \hat{\boldJ}_{\boldQ}\rangle={}^{t}(0,\pm ib, c)$ with $c\neq b$.  
The situation is similar even for the two-directional-screw state. 
This problem exists even for a small value of $J_{\textrm{odd}}$, 
for which the $J_{\textrm{odd}}$ terms act 
as the weak perturbation against the $J_{0}$ term and the $D$ terms. 
Such weak perturbation will not break the stability of the one-directional-screw state 
at least in a non-frustrated system. 
Thus, 
this result highlights 
the limit of the hard-pseudospin approximation 
in discussing the stability of the screw states 
in the presence of the anisotropic exchange interactions. 
Namely, for such discussions, 
we need to take into account the quantum fluctuations, 
which cause softness of the pseudospins. 
Further details about the meanings of this result 
are discussed in Sec. IV. 
Since this work is the first step towards a satisfactory understanding of 
the spin-orbital-coupled chirality in a non-frustrated Mott insulator 
with the strong SOC, 
the analysis including the quantum fluctuations 
is a future work. 

\section{Discussion}
I first discuss the meanings of the limit of the hard-pseudospin approximation 
in detail. 
The key to the limit is a conflict between the states stabilized 
by the DM-type and the Kitaev-type interactions. 
The DM-type interactions stabilize the states 
in which some components of $\langle \hat{J}_{\boldQ}^{\alpha}\rangle$ 
(e.g., $\langle \hat{J}_{\boldQ}^{x}\rangle$ and $\langle \hat{J}_{\boldQ}^{z}\rangle$) 
are mixed under a certain condition (e.g., $|\langle \hat{J}_{\boldQ}^{x}\rangle|
=|\langle \hat{J}_{\boldQ}^{z}\rangle|$). 
On the other hand, 
the Kitaev-type interactions stabilize 
the states in which 
one component is different from the others 
(e.g., for the Kitaev-type interactions of the $z$ component, 
$|\langle \hat{J}_{\boldQ}^{z}\rangle|\neq
|\langle \hat{J}_{\boldQ}^{x}\rangle|, |\langle \hat{J}_{\boldQ}^{y}\rangle|$). 
Thus, 
the states stabilized by the DM-type and the Kitaev-type interactions 
are generally incompatible 
within the hard-pseudospin approximation. 
This property may hold in other systems with the strong SOC, 
where the DM-type and the Kitaev-type interactions appear. 
Then, 
this property has difficulty in calculating 
the pseudospin-wave dispersions 
because 
the pseudospin-wave dispersions are usually calculated 
by considering the quantum fluctuations around the most stable state 
in the hard pseudospin approximation. 
Since the hard-pseudospin approximation 
is used in not only the mean-field approximation 
but also the Luttinger-Tisza method~\cite{Luttinger-Tisza} 
and the classical Monte Carlo calculation, 
which are frequently used in the theoretical studies 
for systems with the strong SOC, 
the result of the limit of the hard-pseudospin approximation 
provides an important step 
for research of the systems with the strong SOC. 

Then, 
we discuss the effects of the tetragonal crystal field 
and the $J_{\textrm{eff}}=\frac{3}{2}$ states. 
As we will see below, 
the treatment of those in this paper is appropriate 
for qualitative analyses. 
In particular, 
the treatment is sufficient to clarify 
the main issue, i.e., 
whether not only the spin but also the orbital 
acquires the chirality for the strong SOC 
in the situation in which the spin acquires the chirality for the weak SOC. 

We begin with the effect of the tetragonal crystal field, $\Delta_{\textrm{tetra}}$. 
Since we consider a $214$-type perovskite oxide, such as Sr$_{2}$IrO$_{4}$, 
$\Delta_{\textrm{tetra}}$ becomes finite; 
$\Delta_{\textrm{tetra}}$ splits the $t_{2g}$ orbitals 
into the $d_{xy}$ orbital and the degenerate $d_{xz}$ and $d_{yz}$ orbitals.  
To show the validity of neglecting its effect for qualitative analyses, 
let us consider two limiting cases, 
(i) $\lambda_{LS}\ll \Delta_{\textrm{tetra}}$ and  
(ii) $\lambda_{LS}\gg \Delta_{\textrm{tetra}}$; 
the limiting cases are sufficient for qualitative analyses, 
while the nonlimiting cases are necessary for quantitative analyses. 
Case (i) is inappropriate to analyze the magnetic properties 
in the presence of the formation of the spin-orbital-coupled degree of freedom 
because the ground state in case (i) 
is the state that one hole occupies either the $d_{xy}$ orbital or 
the degenerate $d_{xz}$ and $d_{yz}$ orbitals, depending on the sign of $\Delta_{\textrm{tetra}}$. 
On the other hand, 
in case (ii), 
we can analyze the effect of $\Delta_{\textrm{tetra}}$ 
on the magnetic properties for the $J_{\textrm{eff}}$ states. 
Since we consider $\lambda_{LS}\gg \Delta_{\textrm{tetra}}$ 
in $U,U^{\prime}\gg \lambda_{LS}\gg 
|t_{ab;\boldi\boldj}^{(\textrm{even})}|, 
|t_{ab;\boldi\boldj}^{(\textrm{odd})}|$, 
the effect of $\Delta_{\textrm{tetra}}$ is negligible compared with the effect of $U$.
 
We turn to the effect of the $J_{\textrm{eff}}=\frac{3}{2}$ states. 
Due to the non-perturbative treatment of the SOC, 
the $t_{2g}$ orbitals are split into the $J_{\textrm{eff}}=\frac{1}{2}$ states 
and the $J_{\textrm{eff}}=\frac{3}{2}$ states. 
In the ground state for the $(t_{2g})^{5}$-electron configuration, 
four electrons per site occupy the $J_{\textrm{eff}}=\frac{3}{2}$ states 
and one electron per site occupies the $J_{\textrm{eff}}=\frac{1}{2}$ states; 
this is equivalent to the configuration in which 
one hole per site occupies the $J_{\textrm{eff}}=\frac{1}{2}$ states. 
In this configuration, 
the $J_{\textrm{eff}}=\frac{3}{2}$ states do not affect the initial and final states 
of the perturbation calculations for the ground state 
because the initial and final states are the lowest-energy states 
for the non-perturbative Hamiltonian. 
On the other hand, 
the $J_{\textrm{eff}}=\frac{3}{2}$ states affect the intermediate states 
of the perturbation calculations. 
This effect is taken into account in our calculations. 
This is because 
we consider $U\gg \lambda_{LS}$, 
in which the intermediate states are approximately given by the two-hole's eigenstates 
for $\hat{H}_{\textrm{int}}$ [see Eq. (\ref{eq:Heff-12-simpler})], 
and because those eigenstates include two-hole's states 
not only for the $J_{\textrm{eff}}=\frac{1}{2}$ states 
but also for the $J_{\textrm{eff}}=\frac{3}{2}$ states. 
The above treatment of the $J_{\textrm{eff}}=\frac{3}{2}$ states 
is sufficient for qualitative analyses 
of the magnetic properties, 
while for the quantitative analysis, 
we may consider the superexchange interactions 
derived in the second-lowest-energy states 
in which three electrons occupy the $J_{\textrm{eff}}=\frac{3}{2}$ states 
and two electrons occupy the $J_{\textrm{eff}}=\frac{1}{2}$ states. 
 
Finally, 
let us apply the present theory to the quasi-two-dimensional insulating iridates 
near the $[001]$-surface of Sr$_{2}$IrO$_{4}$ 
and the interface between Sr$_{2}$IrO$_{4}$ and Sr$_{3}$Ir$_{2}$O$_{7}$. 
First, 
the effects of the IS breaking near the surface and the interface 
are described by $\hat{H}_{\textrm{odd}}$ 
because $\hat{H}_{\textrm{odd}}$ can describe the effects of $ab$-plane's IS breaking 
on the quasi-two-dimensional $t_{2g}$-orbital systems 
(e.g., the Ru oxides~\cite{Yanase-ISB,Mizoguchi-SHE} and Ti oxides~\cite{Ti-Yanase}). 
In addition, 
the $(\pi,\pi)$ antiferromagnetism~\cite{Ir214-weakFM-exp,Ir214-inplaneAF} in Sr$_{2}$IrO$_{4}$ 
can be understood within the mean-field approximation 
for the superexchange interactions derived from $\hat{H}$ for $t_{\textrm{odd}}=0$. 
Although the $(\pi,\pi)$-antiferromagnetic state becomes 
most stable for the realistic parameters, 
$J_{\textrm{H}}>0$ and $|t_{11}|>|t_{33}|$ are necessary 
to understand the difference between the in-plane and the out-of-plane 
alignments of the $(\pi,\pi)$-antiferromagnetic moments. 
This is because 
the main terms stabilizing the in-plane and the out-of-plane alignments for $t_{\textrm{odd}}=0$
are 
$\frac{8}{9}\frac{(t_{11})^{2}J_{\textrm{H}}}{(U^{\prime})^{2}+(J_{\textrm{H}})^{2}}
[\textstyle\sum_{\langle \boldi,\boldj \rangle_{x}}\hat{J}_{\boldi}^{y}\hat{J}_{\boldj}^{y}
+\textstyle\sum_{\langle \boldi,\boldj \rangle_{y}}\hat{J}_{\boldi}^{x}\hat{J}_{\boldj}^{x}]
+\frac{16}{9}\frac{t_{11}t_{33}J_{\textrm{H}}}{(U^{\prime})^{2}+(J_{\textrm{H}})^{2}}
[\textstyle\sum_{\langle \boldi,\boldj \rangle_{x}}\hat{J}_{\boldi}^{x}\hat{J}_{\boldj}^{x}
+\textstyle\sum_{\langle \boldi,\boldj \rangle_{y}}\hat{J}_{\boldi}^{y}\hat{J}_{\boldj}^{y}]$
and 
$\frac{8}{9}\frac{(t_{33})^{2}J_{\textrm{H}}}{(U^{\prime})^{2}+(J_{\textrm{H}})^{2}}
\textstyle\sum_{\langle \boldi,\boldj \rangle}\hat{J}_{\boldi}^{z}\hat{J}_{\boldj}^{z}$, 
respectively [see Eqs. (\ref{eq:Heff-12-x}) and (\ref{eq:Heff-12-y})], 
and because the former terms 
are dominant for $|t_{11}|>|t_{33}|$. 
However, 
even in the presence of the anisotropies for $J_{\textrm{H}}>0$, 
the screw states may be more stable 
than the $(\pi,\pi)$-antiferromagnetic state 
by introducing $ab$-plane's IS breaking 
if the Heisenberg-type interactions between the $z$ components remain finite. 
This is because 
the finite Heisenberg-type interactions between the $z$ components 
and the $x/y$ components are necessary to 
stabilize the screw states by using the DM-type interactions 
due to $ab$-plane's IS breaking. 
Then, 
the effects of the rotation~\cite{Ir214-RotAngle} of IrO$_{6}$ octahedra 
for the small angle $(\simeq 11^{\circ})$ 
will not qualitatively change 
the emergence of the spin-orbital-coupled vector chirality. 
This is because its most drastic effect 
is to induce the small canted angle of 
the in-plane aligned moments of Sr$_{2}$IrO$_{4}$~\cite{Ir214-weakFM-theory}, 
and because 
the coefficient of the $D$ terms is made to be larger 
than the coefficients of the rotation-induced exchange interactions~\cite{Ir214-weakFM-theory} 
(which are small for the small angle) 
by tuning the value of $t_{\textrm{odd}}$ 
with keeping the rotation angle small. 
Thus, 
the candidates for the spin-orbital-coupled vector chirality are 
the quasi-two-dimensional insulating iridates 
near the $[001]$-surface of Sr$_{2}$IrO$_{4}$ 
and the interface between Sr$_{2}$IrO$_{4}$ and Sr$_{3}$Ir$_{2}$O$_{7}$. 

\section{Summary}
In summary, 
I have studied the effects of the IS broken 
near an $ab$ plane of a quasi-two-dimensional insulating iridate 
using the low-energy effective Hamiltonian of the superexchange interactions. 
I find that 
the DM-type interactions, induced by the IS breaking, 
cause the spin-orbital-coupled vector chirality 
as a result of stabilizing the screw state 
compared with the $(\pi,\pi)$-antiferromagnetic state. 
I also find that 
in the presence of both the DM-type and the Kitaev-type interactions, 
the hard-pseudospin approximation becomes inappropriate 
to analyze the stability of the screw state. 
Then, 
I discuss the effects of the tetragonal crystal field and the $J_{\textrm{eff}}=\frac{3}{2}$ states, 
and show the validity of their treatment for qualitative analyses. 
I finally argue that 
the candidates for realizing the spin-orbital-coupled vector chirality 
are the iridates near 
the $[001]$ surface of Sr$_{2}$IrO$_{4}$ and 
the interface between Sr$_{2}$IrO$_{4}$ and Sr$_{3}$Ir$_{2}$O$_{7}$. 
The finding of the spin-orbital-coupled vector chirality 
provides a new possibility 
of utilizing the chirality 
of the anisotropic spatial distribution of electrons/holes 
in a non-frustrated Mott insulator with the strong SOC 
by introducing the IS breaking.

\begin{acknowledgments}
The author thanks T. Mizoguchi 
for useful comments about the previous theoretical studies in the frustrated iridates. 
For the numerical calculations, 
the author used the facilities of 
the Super Computer Center, 
the Institute for Solid State Physics, 
the University of Tokyo. 
\end{acknowledgments}

\appendix

\section{Derivation of Eq. (\ref{eq:Heff})}

In this appendix, 
I derive Eq. (\ref{eq:Heff}) 
by calculating the superexchange interactions in the Mott insulator 
for $\hat{H}$. 
This derivation is the extension of the formulation~\cite{Superex-titanates} 
for a $t_{2g}$-orbital Hubbard model without the SOC 
to the case with the SOC. 
The treatment of the SOC is similar to that 
for Refs. \onlinecite{hyperkagomeIr-Mizoguchi} and \onlinecite{honeycombIr-zigzag}. 

In this derivation, 
we use three assumptions, 
resulting in the condition $U,U^{\prime}\gg \lambda_{LS} 
\gg |t_{ab;\boldi\boldj}^{(\textrm{even})}|, |t_{ab;\boldi\boldj}^{(\textrm{odd})}|$. 
We first assume that 
the hopping integrals of $\hat{H}_{\textrm{even}}$ and $\hat{H}_{\textrm{odd}}$, 
$t_{ab;\boldi\boldj}^{(\textrm{even})}$ and $t_{ab;\boldi\boldj}^{(\textrm{odd})}$, 
are smaller than $U$ and $U^{\prime}$. 
Thus, 
we can treat the effects of $\hat{H}_{\textrm{even}}$ and $\hat{H}_{\textrm{odd}}$ 
as the second-order perturbation against the interaction terms. 
Also, 
we assume that 
the SOC, $\lambda_{LS}$, 
is larger than $|t_{ab;\boldi\boldj}^{(\textrm{even})}|$ and $|t_{ab;\boldi\boldj}^{(\textrm{odd})}|$; 
as a result, 
the excitations from the $J_{\textrm{eff}}=\frac{1}{2}$ states 
to the $J_{\textrm{eff}}=\frac{3}{2}$ states 
are negligible. 
Thus, 
the nonperturbed states of two sites ($\boldi=\boldone, \boldtwo$) 
are given by the products of the $J^{z}=\pm\frac{1}{2}$ states~\cite{Rigand-text}, 
i.e. $|+,+ \rangle =|\boldone, + \rangle |\boldtwo, + \rangle$, 
$|+,- \rangle =|\boldone, + \rangle |\boldtwo, - \rangle$,
$|-,+ \rangle =|\boldone, - \rangle |\boldtwo, + \rangle$, 
and $|-,- \rangle =|\boldone, - \rangle |\boldtwo, - \rangle$ with
\begin{align}
&|\boldi, + \rangle 
=
\frac{1}{\sqrt{3}}
(i\hat{c}^{\dagger}_{\boldi d_{xz}\downarrow}
+\hat{c}^{\dagger}_{\boldi d_{yz}\downarrow}
+\hat{c}^{\dagger}_{\boldi d_{xy}\uparrow})
|0 \rangle,\label{eq:Jzp}\\
&|\boldi, - \rangle 
=
\frac{1}{\sqrt{3}}
(i\hat{c}^{\dagger}_{\boldi d_{xz}\uparrow}
-\hat{c}^{\dagger}_{\boldi d_{yz}\uparrow}
+\hat{c}^{\dagger}_{\boldi d_{xy}\downarrow})
|0 \rangle.\label{eq:Jzm} 
\end{align}
Moreover, for simplicity of the formulation, 
we assume that 
$\lambda_{LS}$ is smaller than $U$ and $U^{\prime}$. 
Because of this assumption, 
we can neglect the effects of $\lambda_{\textrm{LS}}$ on the energy of the intermediate states 
of the second-order perturbation, 
i.e. $\frac{1}{E_{0}-\hat{H}_{\textrm{int}}-\hat{H}_{LS}}\sim 
\frac{1}{-\hat{H}_{\textrm{int}}}$. 
Note that 
since $\hat{H}_{\textrm{int}}$ is rewritten in terms of the irreducible representations 
for the two-hole states per site [see Eq. (\ref{eq:Hint-irrep})], 
the condition $U,U^{\prime}\gg \lambda_{LS}$ implies that 
the energies of all the irreducible representations, 
which include either $U$ or $U^{\prime}$, are larger 
than $\lambda_{LS}$, i.e. $U_{\Gamma}\gg \lambda_{LS}$. 

Under the condition 
$U,U^{\prime}\gg \lambda_{LS} 
\gg |t_{ab;\boldi\boldj}^{(\textrm{even})}|, |t_{ab;\boldi\boldj}^{(\textrm{odd})}|$, 
we derive the Hamiltonian of the superexchange interactions 
between the two neighboring sites, $(\hat{H}_{\textrm{eff}})_{\boldone \boldtwo}$, from 
\begin{align}
(\hat{H}_{\textrm{eff}})_{\boldone \boldtwo}
=& 
\langle f|(\hat{H}_{\textrm{even}}+\hat{H}_{\textrm{odd}})
\frac{1}{-\hat{H}_{\textrm{int}}}
(\hat{H}_{\textrm{even}}+\hat{H}_{\textrm{odd}})|i\rangle \notag\\
&\times
|f\rangle\langle i|,\label{eq:Heff-12}
\end{align}
with $\{|i\rangle,|f\rangle\}
=\{|+,+ \rangle,|+,- \rangle,|-,+ \rangle,|-,- \rangle\}$. 
For example, 
for $|i\rangle=|+,+\rangle$ and $|f\rangle=|+,+\rangle$, 
$|f\rangle\langle i|$ is given by 
the operator $(\frac{1}{2}\hat{n}_{\boldone}-\hat{J}_{\boldone}^{z})
(\frac{1}{2}\hat{n}_{\boldtwo}-\hat{J}_{\boldtwo}^{z})$; 
for $|i\rangle=|+,-\rangle$ and $|f\rangle=|-,+\rangle$, 
$|f\rangle\langle i|$ is given by 
the operator $\hat{J}_{\boldone}^{-}\hat{J}_{\boldtwo}^{+}$. 

For easy treatment of $\hat{H}_{\textrm{int}}$ in Eq. (\ref{eq:Heff-12}), 
we rewrite $\hat{H}_{\textrm{int}}$ in terms of 
the irreducible representations~\cite{Superex-titanates} for the two-hole states, 
the intermediate states in the second-order perturbation: 
\begin{align}
\hat{H}_{\textrm{int}}
=\sum\limits_{\boldi}\sum\limits_{\Gamma}\sum\limits_{g_{\Gamma}}
U_{\Gamma}|\boldi;\Gamma,g_{\Gamma}\rangle 
\langle \boldi;\Gamma,g_{\Gamma}|,\label{eq:Hint-irrep}
\end{align}
where $\Gamma$ denotes the irreducible representations, 
and $g_{\Gamma}$ denotes the degeneracy. 
For the two-hole states of the $t_{2g}$-orbital Hubbard model, 
there are four kinds of $U_{\Gamma}$: 
\begin{align}
&U_{A_{1}}=U+2J^{\prime},\label{eq:U-Irrep1}\\
&U_{E}=U-J^{\prime},\label{eq:U-Irrep2}\\
&U_{T_{1}}=U^{\prime}-J_{\textrm{H}},\label{eq:U-Irrep3}\\
&U_{T_{2}}=U^{\prime}+J_{\textrm{H}};\label{eq:U-Irrep4}
\end{align}
and $|\boldi;\Gamma,g_{\Gamma}\rangle$ are
\begin{align}
&|\boldi;A_{1} \rangle 
=
\frac{1}{\sqrt{3}}
(\hat{c}^{\dagger}_{\boldi d_{xz}\uparrow}\hat{c}^{\dagger}_{\boldi d_{xz}\downarrow}
+\hat{c}^{\dagger}_{\boldi d_{yz}\uparrow}\hat{c}^{\dagger}_{\boldi d_{yz}\downarrow}\notag\\
&\ \ \ \ \ \ \ \ \ \ \ +\hat{c}^{\dagger}_{\boldi d_{xy}\uparrow}\hat{c}^{\dagger}_{\boldi d_{xy}\downarrow})
|0 \rangle,\label{eq:basis-Irrep1}\\
&|\boldi;E,u \rangle 
=
\sqrt{\frac{2}{3}}
(-\hat{c}^{\dagger}_{\boldi d_{xz}\uparrow}\hat{c}^{\dagger}_{\boldi d_{xz}\downarrow}
+\frac{1}{2}\hat{c}^{\dagger}_{\boldi d_{yz}\uparrow}\hat{c}^{\dagger}_{\boldi d_{yz}\downarrow}\notag\\
&\ \ \ \ \ \ \ \ \ \ \ +\frac{1}{2}\hat{c}^{\dagger}_{\boldi d_{xy}\uparrow}\hat{c}^{\dagger}_{\boldi d_{xy}\downarrow})
|0 \rangle,\label{eq:basis-Irrep2}\\
&|\boldi;E,v \rangle 
=
\frac{1}{\sqrt{2}}
(\hat{c}^{\dagger}_{\boldi d_{yz}\uparrow}\hat{c}^{\dagger}_{\boldi d_{yz}\downarrow}
-\hat{c}^{\dagger}_{\boldi d_{xy}\uparrow}\hat{c}^{\dagger}_{\boldi d_{xy}\downarrow})
|0 \rangle,\label{eq:basis-Irrep3}\\
&|\boldi;T_{1},\zeta_{+} \rangle 
=
\hat{c}^{\dagger}_{\boldi d_{xz}\uparrow}\hat{c}^{\dagger}_{\boldi d_{yz}\uparrow}
|0 \rangle,\label{eq:basis-Irrep4}\\
&|\boldi;T_{1},\zeta_{-} \rangle 
=
\hat{c}^{\dagger}_{\boldi d_{xz}\downarrow}\hat{c}^{\dagger}_{\boldi d_{yz}\downarrow}
|0 \rangle,\label{eq:basis-Irrep5}\\
&|\boldi;T_{1},\zeta_{0} \rangle 
=
\frac{1}{\sqrt{2}}
(\hat{c}^{\dagger}_{\boldi d_{xz}\uparrow}\hat{c}^{\dagger}_{\boldi d_{yz}\downarrow}
+\hat{c}^{\dagger}_{\boldi d_{xz}\downarrow}\hat{c}^{\dagger}_{\boldi d_{yz}\uparrow})
|0 \rangle,\label{eq:basis-Irrep6}\\
&|\boldi;T_{2},\zeta_{0} \rangle 
=
\frac{1}{\sqrt{2}}
(\hat{c}^{\dagger}_{\boldi d_{xz}\uparrow}\hat{c}^{\dagger}_{\boldi d_{yz}\downarrow}
-\hat{c}^{\dagger}_{\boldi d_{xz}\downarrow}\hat{c}^{\dagger}_{\boldi d_{yz}\uparrow})
|0 \rangle,\label{eq:basis-Irrep7}\\
&|\boldi;T_{1},\xi_{+} \rangle 
=
\hat{c}^{\dagger}_{\boldi d_{xz}\uparrow}\hat{c}^{\dagger}_{\boldi d_{xy}\uparrow}
|0 \rangle,\label{eq:basis-Irrep8}\\
&|\boldi;T_{1},\xi_{-} \rangle 
=
\hat{c}^{\dagger}_{\boldi d_{xz}\downarrow}\hat{c}^{\dagger}_{\boldi d_{xy}\downarrow}
|0 \rangle,\label{eq:basis-Irrep9}\\
&|\boldi;T_{1},\xi_{0} \rangle 
=
\frac{1}{\sqrt{2}}
(\hat{c}^{\dagger}_{\boldi d_{xz}\uparrow}\hat{c}^{\dagger}_{\boldi d_{xy}\downarrow}
+\hat{c}^{\dagger}_{\boldi d_{xz}\downarrow}\hat{c}^{\dagger}_{\boldi d_{xy}\uparrow})
|0 \rangle,\label{eq:basis-Irrep10}\\
&|\boldi;T_{2},\xi_{0} \rangle 
=
\frac{1}{\sqrt{2}}
(\hat{c}^{\dagger}_{\boldi d_{xz}\uparrow}\hat{c}^{\dagger}_{\boldi d_{xy}\downarrow}
-\hat{c}^{\dagger}_{\boldi d_{xz}\downarrow}\hat{c}^{\dagger}_{\boldi d_{xy}\uparrow})
|0 \rangle,\label{eq:basis-Irrep11}\\
&|\boldi;T_{1},\eta_{+} \rangle 
=
\hat{c}^{\dagger}_{\boldi d_{yz}\uparrow}\hat{c}^{\dagger}_{\boldi d_{xy}\uparrow}
|0 \rangle,\label{eq:basis-Irrep12}\\
&|\boldi;T_{1},\eta_{-} \rangle 
=
\hat{c}^{\dagger}_{\boldi d_{yz}\downarrow}\hat{c}^{\dagger}_{\boldi d_{xy}\downarrow}
|0 \rangle,\label{eq:basis-Irrep13}\\
&|\boldi;T_{1},\eta_{0} \rangle 
=
\frac{1}{\sqrt{2}}
(\hat{c}^{\dagger}_{\boldi d_{yz}\uparrow}\hat{c}^{\dagger}_{\boldi d_{xy}\downarrow}
+\hat{c}^{\dagger}_{\boldi d_{yz}\downarrow}\hat{c}^{\dagger}_{\boldi d_{xy}\uparrow})
|0 \rangle,\label{eq:basis-Irrep14}\\
&|\boldi;T_{2},\eta_{0} \rangle 
=
\frac{1}{\sqrt{2}}
(\hat{c}^{\dagger}_{\boldi d_{yz}\uparrow}\hat{c}^{\dagger}_{\boldi d_{xy}\downarrow}
-\hat{c}^{\dagger}_{\boldi d_{yz}\downarrow}\hat{c}^{\dagger}_{\boldi d_{yx}\uparrow})
|0 \rangle.\label{eq:basis-Irrep15}
\end{align}
By substituting Eq. (\ref{eq:Hint-irrep}) into Eq. (\ref{eq:Heff-12}), 
Eq. (\ref{eq:Heff-12}) becomes 
\begin{align}
&(\hat{H}_{\textrm{eff}})_{\boldone \boldtwo}
= 
\sum\limits_{\boldi=\boldone,\boldtwo}
\sum\limits_{\Gamma}
\sum\limits_{g_{\Gamma}}
\langle f|(\hat{H}_{\textrm{even}}+\hat{H}_{\textrm{odd}})|\boldi;\Gamma,g_{\Gamma}\rangle\notag\\
&\times \frac{1}{-U_{\Gamma}}
\langle \boldi;\Gamma,g_{\Gamma} |
(\hat{H}_{\textrm{even}}+\hat{H}_{\textrm{odd}})|i\rangle 
|f\rangle\langle i|.\label{eq:Heff-12-simpler}
\end{align}
Thus, 
the remaining tasks are to calculate the right-hand side of Eq. (\ref{eq:Heff-12-simpler}) 
for $\boldtwo-\boldone=(1,0)$, $(0,1)$, $(1,1)$, and $(1,-1)$. 
Those components are sufficient to derive the superexchange interactions for $\hat{H}$ 
since all the components in the model considered are categorized into the terms along $x$, $y$, 
$[110]$, and $[1\bar{1}0]$ directions. 

Let us first derive the terms of $(\hat{H}_{\textrm{eff}})_{\boldone \boldtwo}$ 
for $\boldtwo-\boldone =(1,0)$. 
In this case, 
the finite terms of $(\hat{H}_{\textrm{even}}+\hat{H}_{\textrm{odd}})$ 
come from the finite hopping integrals along the $x$ direction: 
for $\hat{H}_{\textrm{even}}$, 
the hopping integral between the $d_{xz}$ orbitals at $\boldone$ and $\boldtwo$, $-t_{11}$, 
and the hopping integral between the $d_{xy}$ orbitals, $-t_{33}$; 
for $\hat{H}_{\textrm{odd}}$, 
the hopping integral between the $d_{yz}$ orbital at $\boldone$ 
and the $d_{xy}$ orbital at $\boldtwo$, $-t_{\textrm{odd}}$, 
and the hopping integral between the $d_{xy}$ orbital at $\boldone$ 
and the $d_{yz}$ orbital at $\boldtwo$, $+t_{\textrm{odd}}$. 
By applying one of those hopping terms to $|\textrm{i}\rangle$, 
one of the four degenerate states (i.e., $|+,+\rangle$, $|+,-\rangle$, 
$|-,+\rangle$, and $|-,-\rangle$), 
and using Eqs. (\ref{eq:basis-Irrep1}){--}(\ref{eq:basis-Irrep15}), 
we obtain the finite terms of 
$\langle \boldi;\Gamma,g_{\Gamma} |(\hat{H}_{\textrm{even}}+\hat{H}_{\textrm{odd}})|i\rangle$ 
for $\boldi=\boldone$ or $\boldtwo$. 
We similarly obtain the finite terms of 
$\langle f|(\hat{H}_{\textrm{even}}+\hat{H}_{\textrm{odd}})|\boldi;\Gamma,g_{\Gamma}\rangle$ 
for $\boldi=\boldone$ or $\boldtwo$. 
By combining those results with Eq. (\ref{eq:Heff-12-simpler}), 
the superexchange interactions for $\boldtwo-\boldone=(1,0)$ are given by
\begin{align}
&(\hat{H}_{\textrm{eff}})_{\boldone \boldtwo}
=
-\frac{4}{27}
(\frac{t_{11}^{2}}{U+2J^{\prime}}+\frac{2t_{11}^{2}}{U-J^{\prime}})
(\frac{1}{4}\hat{n}_{\boldone}\hat{n}_{\boldtwo}
-\hat{\boldJ}_{\boldone}\cdot \hat{\boldJ}_{\boldtwo})\notag\\
&-\frac{4}{9}
\frac{t_{11}^{2}}{U^{\prime}-J_{\textrm{H}}}
(\frac{3}{4}\hat{n}_{\boldone}\hat{n}_{\boldtwo}
-\hat{J}^{y}_{\boldone}\hat{J}^{y}_{\boldtwo})
-\frac{4}{9}
\frac{t_{11}^{2}}{U^{\prime}+J_{\textrm{H}}}
(\frac{1}{4}\hat{n}_{\boldone}\hat{n}_{\boldtwo}
+\hat{J}^{y}_{\boldone}\hat{J}^{y}_{\boldtwo})\notag\\
&-\frac{4}{27}
(\frac{t_{33}^{2}}{U+2J^{\prime}}+\frac{2t_{33}^{2}}{U-J^{\prime}})
(\frac{1}{4}\hat{n}_{\boldone}\hat{n}_{\boldtwo}
-\hat{\boldJ}_{\boldone}\cdot \hat{\boldJ}_{\boldtwo})\notag\\
&-\frac{4}{9}
\frac{t_{33}^{2}}{U^{\prime}-J_{\textrm{H}}}
(\frac{3}{4}\hat{n}_{\boldone}\hat{n}_{\boldtwo}
-\hat{J}^{z}_{\boldone}\hat{J}^{z}_{\boldtwo})
-\frac{4}{9}
\frac{t_{33}^{2}}{U^{\prime}+J_{\textrm{H}}}
(\frac{1}{4}\hat{n}_{\boldone}\hat{n}_{\boldtwo}
+\hat{J}^{z}_{\boldone}\hat{J}^{z}_{\boldtwo})\notag\\
&-\frac{8}{27}
(\frac{t_{11}t_{33}}{U+2J^{\prime}}-\frac{t_{11}t_{33}}{U-J^{\prime}})
(\frac{1}{4}\hat{n}_{\boldone}\hat{n}_{\boldtwo}
-\hat{\boldJ}_{\boldone}\cdot \hat{\boldJ}_{\boldtwo})\notag\\
&+\frac{4}{9}
\frac{t_{11}t_{33}}{U^{\prime}-J_{\textrm{H}}}
(\frac{1}{4}\hat{n}_{\boldone}\hat{n}_{\boldtwo}
+\hat{\boldJ}_{\boldone}\cdot \hat{\boldJ}_{\boldtwo}
+2\hat{J}^{x}_{\boldone}\hat{J}^{x}_{\boldtwo})\notag\\
&+\frac{4}{9}
\frac{t_{11}t_{33}}{U^{\prime}+J_{\textrm{H}}}
(\frac{1}{4}\hat{n}_{\boldone}\hat{n}_{\boldtwo}
+\hat{\boldJ}_{\boldone}\cdot \hat{\boldJ}_{\boldtwo}
-2\hat{J}^{x}_{\boldone}\hat{J}^{x}_{\boldtwo})\notag\\
&-\frac{8}{27}
(\frac{2t_{\textrm{odd}}^{2}}{U+2J^{\prime}}+\frac{t_{\textrm{odd}}^{2}}{U-J^{\prime}})
(\frac{1}{4}\hat{n}_{\boldone}\hat{n}_{\boldtwo}
+\hat{\boldJ}_{\boldone}\cdot \hat{\boldJ}_{\boldtwo}
-2\hat{J}^{y}_{\boldone}\hat{J}^{y}_{\boldtwo})\notag\\
&-\frac{4}{9}
\frac{t_{\textrm{odd}}^{2}}{U^{\prime}-J_{\textrm{H}}}
(\frac{5}{4}\hat{n}_{\boldone}\hat{n}_{\boldtwo}
+2\hat{\boldJ}_{\boldone}\cdot \hat{\boldJ}_{\boldtwo}
-5\hat{J}^{y}_{\boldone}\hat{J}^{y}_{\boldtwo})\notag\\
&-\frac{4}{9}
\frac{t_{\textrm{odd}}^{2}}{U^{\prime}+J_{\textrm{H}}}
(\frac{1}{4}\hat{n}_{\boldone}\hat{n}_{\boldtwo}
+\hat{J}^{y}_{\boldone}\hat{J}^{y}_{\boldtwo})\notag\\
&-\frac{16}{27}
(\frac{t_{11}t_{\textrm{odd}}}{U+2J^{\prime}}-\frac{t_{11}t_{\textrm{odd}}}{U-J^{\prime}})
(\hat{J}^{x}_{\boldone}\hat{J}^{z}_{\boldtwo}-\hat{J}^{z}_{\boldone}\hat{J}^{x}_{\boldtwo})\notag\\
&-\frac{16}{9}
\frac{t_{11}t_{\textrm{odd}}}{U^{\prime}-J_{\textrm{H}}}
(\hat{J}^{x}_{\boldone}\hat{J}^{z}_{\boldtwo}-\hat{J}^{z}_{\boldone}\hat{J}^{x}_{\boldtwo})\notag\\
&-\frac{16}{27}
(\frac{t_{33}t_{\textrm{odd}}}{U+2J^{\prime}}+\frac{1}{2}\frac{t_{33}t_{\textrm{odd}}}{U-J^{\prime}})
(\hat{J}^{x}_{\boldone}\hat{J}^{z}_{\boldtwo}-\hat{J}^{z}_{\boldone}\hat{J}^{x}_{\boldtwo})\notag\\
&-\frac{8}{9}
\frac{t_{33}t_{\textrm{odd}}}{U^{\prime}-J_{\textrm{H}}}
(\hat{J}^{x}_{\boldone}\hat{J}^{z}_{\boldtwo}-\hat{J}^{z}_{\boldone}\hat{J}^{x}_{\boldtwo}).\label{eq:Heff-12-x}
\end{align}
In the above derivation, 
we have used the relations, 
$\hat{J}_{\boldone}^{+}\hat{J}_{\boldtwo}^{-}+\hat{J}_{\boldone}^{-}\hat{J}_{\boldtwo}^{+}
=2(\hat{J}_{\boldone}^{x}\hat{J}_{\boldtwo}^{x}+\hat{J}_{\boldone}^{y}\hat{J}_{\boldtwo}^{y})$, 
$\hat{J}_{\boldone}^{+}\hat{J}_{\boldtwo}^{+}+\hat{J}_{\boldone}^{-}\hat{J}_{\boldtwo}^{-}
=2(\hat{J}_{\boldone}^{x}\hat{J}_{\boldtwo}^{x}-\hat{J}_{\boldone}^{y}\hat{J}_{\boldtwo}^{y})$, 
and 
$\hat{J}_{\boldj}^{+}+\hat{J}_{\boldj}^{-}=2\hat{J}_{\boldj}^{x}$. 
If we set $J_{\textrm{H}}=0$, $J^{\prime}=0$, and $U^{\prime}=U$ in Eq. (\ref{eq:Heff-12-x}), 
we obtain 
\begin{align}
&(\hat{H}_{\textrm{eff}})_{\boldone \boldtwo}
=
-\frac{4}{9}
\frac{t_{11}^{2}}{U}
(\frac{5}{4}\hat{n}_{\boldone}\hat{n}_{\boldtwo}
-\hat{\boldJ}_{\boldone}\cdot \hat{\boldJ}_{\boldtwo})\notag\\
&-\frac{4}{9}
\frac{t_{33}^{2}}{U}
(\frac{5}{4}\hat{n}_{\boldone}\hat{n}_{\boldtwo}
-\hat{\boldJ}_{\boldone}\cdot \hat{\boldJ}_{\boldtwo})
+\frac{8}{9}
\frac{t_{11}t_{33}}{U}
(\frac{1}{4}\hat{n}_{\boldone}\hat{n}_{\boldtwo}
+\hat{\boldJ}_{\boldone}\cdot \hat{\boldJ}_{\boldtwo})\notag\\
&-\frac{16}{9}
\frac{t_{\textrm{odd}}^{2}}{U}
(\frac{1}{2}\hat{n}_{\boldone}\hat{n}_{\boldtwo}
+\hat{\boldJ}_{\boldone}\cdot \hat{\boldJ}_{\boldtwo}
-2\hat{J}^{y}_{\boldone}\hat{J}^{y}_{\boldtwo})\notag\\
&-\frac{16}{9}
\frac{(t_{11}+t_{33})t_{\textrm{odd}}}{U}
(\hat{J}^{x}_{\boldone}\hat{J}^{z}_{\boldtwo}-\hat{J}^{z}_{\boldone}\hat{J}^{x}_{\boldtwo}).
\label{eq:Heff-12-x-J0}
\end{align}

Next, 
we derive the terms of $(\hat{H}_{\textrm{eff}})_{\boldone \boldtwo}$ for $\boldtwo-\boldone=(0,1)$. 
This derivation can be carried out in a similar way for $\boldtwo-\boldone=(1,0)$ 
except the difference in the finite hopping integrals. 
The finite hopping integrals for $\boldtwo-\boldone=(0,1)$ 
come from the hopping integrals of $\hat{H}_{0}$ 
between the $d_{yz}$ orbitals and between the $d_{xy}$ orbitals 
($-t_{11}$ and $-t_{33}$, respectively), 
and the hopping integrals of $\hat{H}_{\textrm{odd}}$ 
between the $d_{xz}$ orbital at $\boldone$ and the $d_{xy}$ orbital at $\boldtwo$ 
and between the $d_{xy}$ orbital at $\boldone$ and the $d_{xz}$ orbital at $\boldtwo$ 
($-t_{\textrm{odd}}$ and $+t_{\textrm{odd}}$, respectively). 
Carrying out similar calculations for $\boldtwo-\boldone=(1,0)$, 
we obtain the superexchange interactions for $\boldtwo-\boldone=(0,1)$:
\begin{align}
&(\hat{H}_{\textrm{eff}})_{\boldone \boldtwo}
=
-\frac{4}{27}
(\frac{t_{11}^{2}}{U+2J^{\prime}}+\frac{2t_{11}^{2}}{U-J^{\prime}})
(\frac{1}{4}\hat{n}_{\boldone}\hat{n}_{\boldtwo}
-\hat{\boldJ}_{\boldone}\cdot \hat{\boldJ}_{\boldtwo})\notag\\
&-\frac{4}{9}
\frac{t_{11}^{2}}{U^{\prime}-J_{\textrm{H}}}
(\frac{3}{4}\hat{n}_{\boldone}\hat{n}_{\boldtwo}
-\hat{J}^{x}_{\boldone}\hat{J}^{x}_{\boldtwo})
-\frac{4}{9}
\frac{t_{11}^{2}}{U^{\prime}+J_{\textrm{H}}}
(\frac{1}{4}\hat{n}_{\boldone}\hat{n}_{\boldtwo}
+\hat{J}^{x}_{\boldone}\hat{J}^{x}_{\boldtwo})\notag\\
&-\frac{4}{27}
(\frac{t_{33}^{2}}{U+2J^{\prime}}+\frac{2t_{33}^{2}}{U-J^{\prime}})
(\frac{1}{4}\hat{n}_{\boldone}\hat{n}_{\boldtwo}
-\hat{\boldJ}_{\boldone}\cdot \hat{\boldJ}_{\boldtwo})\notag\\
&-\frac{4}{9}
\frac{t_{33}^{2}}{U^{\prime}-J_{\textrm{H}}}
(\frac{3}{4}\hat{n}_{\boldone}\hat{n}_{\boldtwo}
-\hat{J}^{z}_{\boldone}\hat{J}^{z}_{\boldtwo})
-\frac{4}{9}
\frac{t_{33}^{2}}{U^{\prime}+J_{\textrm{H}}}
(\frac{1}{4}\hat{n}_{\boldone}\hat{n}_{\boldtwo}
+\hat{J}^{z}_{\boldone}\hat{J}^{z}_{\boldtwo})\notag\\
&-\frac{8}{27}
(\frac{t_{11}t_{33}}{U+2J^{\prime}}-\frac{t_{11}t_{33}}{U-J^{\prime}})
(\frac{1}{4}\hat{n}_{\boldone}\hat{n}_{\boldtwo}
-\hat{\boldJ}_{\boldone}\cdot \hat{\boldJ}_{\boldtwo})\notag\\
&+\frac{4}{9}
\frac{t_{11}t_{33}}{U^{\prime}-J_{\textrm{H}}}
(\frac{1}{4}\hat{n}_{\boldone}\hat{n}_{\boldtwo}
+\hat{\boldJ}_{\boldone}\cdot \hat{\boldJ}_{\boldtwo}
+2\hat{J}^{y}_{\boldone}\hat{J}^{y}_{\boldtwo})\notag\\
&+\frac{4}{9}
\frac{t_{11}t_{33}}{U^{\prime}+J_{\textrm{H}}}
(\frac{1}{4}\hat{n}_{\boldone}\hat{n}_{\boldtwo}
+\hat{\boldJ}_{\boldone}\cdot \hat{\boldJ}_{\boldtwo}
-2\hat{J}^{y}_{\boldone}\hat{J}^{y}_{\boldtwo})\notag\\
&-\frac{8}{27}
(\frac{2t_{\textrm{odd}}^{2}}{U+2J^{\prime}}+\frac{t_{\textrm{odd}}^{2}}{U-J^{\prime}})
(\frac{1}{4}\hat{n}_{\boldone}\hat{n}_{\boldtwo}
+\hat{\boldJ}_{\boldone}\cdot \hat{\boldJ}_{\boldtwo}
-2\hat{J}^{x}_{\boldone}\hat{J}^{x}_{\boldtwo})\notag\\
&-\frac{4}{9}
\frac{t_{\textrm{odd}}^{2}}{U^{\prime}-J_{\textrm{H}}}
(\frac{5}{4}\hat{n}_{\boldone}\hat{n}_{\boldtwo}
+2\hat{\boldJ}_{\boldone}\cdot \hat{\boldJ}_{\boldtwo}
-5\hat{J}^{x}_{\boldone}\hat{J}^{x}_{\boldtwo})\notag\\
&-\frac{4}{9}
\frac{t_{\textrm{odd}}^{2}}{U^{\prime}+J_{\textrm{H}}}
(\frac{1}{4}\hat{n}_{\boldone}\hat{n}_{\boldtwo}
+\hat{J}^{x}_{\boldone}\hat{J}^{x}_{\boldtwo})\notag\\
&-\frac{16}{27}
(\frac{t_{11}t_{\textrm{odd}}}{U+2J^{\prime}}-\frac{t_{11}t_{\textrm{odd}}}{U-J^{\prime}})
(\hat{J}^{y}_{\boldone}\hat{J}^{z}_{\boldtwo}-\hat{J}^{z}_{\boldone}\hat{J}^{y}_{\boldtwo})\notag\\
&-\frac{16}{9}
\frac{t_{11}t_{\textrm{odd}}}{U^{\prime}-J_{\textrm{H}}}
(\hat{J}^{y}_{\boldone}\hat{J}^{z}_{\boldtwo}-\hat{J}^{z}_{\boldone}\hat{J}^{y}_{\boldtwo})\notag\\
&-\frac{16}{27}
(\frac{t_{33}t_{\textrm{odd}}}{U+2J^{\prime}}+\frac{1}{2}\frac{t_{33}t_{\textrm{odd}}}{U-J^{\prime}})
(\hat{J}^{y}_{\boldone}\hat{J}^{z}_{\boldtwo}-\hat{J}^{z}_{\boldone}\hat{J}^{y}_{\boldtwo})\notag\\
&-\frac{8}{9}
\frac{t_{33}t_{\textrm{odd}}}{U^{\prime}-J_{\textrm{H}}}
(\hat{J}^{y}_{\boldone}\hat{J}^{z}_{\boldtwo}-\hat{J}^{z}_{\boldone}\hat{J}^{y}_{\boldtwo}).\label{eq:Heff-12-y}
\end{align} 
Because of the tetragonal symmetry of the system, 
Eq. (\ref{eq:Heff-12-y}) is symbolically equivalent to Eq. (\ref{eq:Heff-12-x}) 
after the replacements $\hat{J}_{\boldj}^{x}\rightarrow \hat{J}_{\boldj}^{y}$ 
and $\hat{J}_{\boldj}^{y}\rightarrow \hat{J}_{\boldj}^{x}$. 
Then, 
from Eqs. (\ref{eq:Heff-12-x}) and (\ref{eq:Heff-12-y}), 
we see that 
the energy difference between 
the in-plane and the out-of-plane alignments of the $(\pi,\pi)$-antiferromagnetic moments 
arises mainly from 
the difference between 
$\frac{8}{9}\frac{(t_{11})^{2}J_{\textrm{H}}}{(U^{\prime})^{2}+(J_{\textrm{H}})^{2}}
[\textstyle\sum_{\langle \boldi,\boldj \rangle_{x}}\hat{J}_{\boldi}^{y}\hat{J}_{\boldj}^{y}
+\textstyle\sum_{\langle \boldi,\boldj \rangle_{y}}\hat{J}_{\boldi}^{x}\hat{J}_{\boldj}^{x}]
+\frac{16}{9}\frac{t_{11}t_{33}J_{\textrm{H}}}{(U^{\prime})^{2}+(J_{\textrm{H}})^{2}}
[\textstyle\sum_{\langle \boldi,\boldj \rangle_{x}}\hat{J}_{\boldi}^{x}\hat{J}_{\boldj}^{x}
+\textstyle\sum_{\langle \boldi,\boldj \rangle_{y}}\hat{J}_{\boldi}^{y}\hat{J}_{\boldj}^{y}]$ 
and $\frac{8}{9}\frac{(t_{33})^{2}J_{\textrm{H}}}{(U^{\prime})^{2}+(J_{\textrm{H}})^{2}}
\textstyle\sum_{\langle \boldi,\boldj \rangle}\hat{J}_{\boldi}^{z}\hat{J}_{\boldj}^{z}$, 
as described in Sec. IV. 
For $J_{\textrm{H}}=0$, $J^{\prime}=0$, and $U^{\prime}=U$, 
Eq. (\ref{eq:Heff-12-y}) reduces to the following equation: 
\begin{align}
&(\hat{H}_{\textrm{eff}})_{\boldone \boldtwo}
=
-\frac{4}{9}
\frac{t_{11}^{2}}{U}
(\frac{5}{4}\hat{n}_{\boldone}\hat{n}_{\boldtwo}
-\hat{\boldJ}_{\boldone}\cdot \hat{\boldJ}_{\boldtwo})\notag\\
&-\frac{4}{9}
\frac{t_{33}^{2}}{U}
(\frac{5}{4}\hat{n}_{\boldone}\hat{n}_{\boldtwo}
-\hat{\boldJ}_{\boldone}\cdot \hat{\boldJ}_{\boldtwo})\notag\\
&+\frac{8}{9}
\frac{t_{11}t_{33}}{U}
(\frac{1}{4}\hat{n}_{\boldone}\hat{n}_{\boldtwo}
+\hat{\boldJ}_{\boldone}\cdot \hat{\boldJ}_{\boldtwo})\notag\\
&-\frac{16}{9}
\frac{t_{\textrm{odd}}^{2}}{U}
(\frac{1}{2}\hat{n}_{\boldone}\hat{n}_{\boldtwo}
+\hat{\boldJ}_{\boldone}\cdot \hat{\boldJ}_{\boldtwo}
-2\hat{J}^{x}_{\boldone}\hat{J}^{x}_{\boldtwo})\notag\\
&-\frac{16}{9}
\frac{(t_{11}+t_{33})t_{\textrm{odd}}}{U}
(\hat{J}^{y}_{\boldone}\hat{J}^{z}_{\boldtwo}-\hat{J}^{z}_{\boldone}\hat{J}^{y}_{\boldtwo}).
\label{eq:Heff-12-y-J0}
\end{align}

Moreover, 
we can derive the terms of $(\hat{H}_{\textrm{eff}})_{\boldone \boldtwo}$ 
for $\boldtwo-\boldone=(1,1)$ and $(1,-1)$. 
Those derivations are simpler than 
the derivations for $\boldtwo-\boldone=(1,0)$ and $(0,1)$ 
because the finite hopping integrals for $\boldtwo-\boldone=(1,\pm 1)$ 
are the hopping integrals between the $d_{xz}$ orbital and the $d_{yz}$ orbital, $\mp t_{12}^{\prime}$, 
and the hopping integral between the $d_{xy}$ orbitals, $-t_{33}^{\prime}$. 
The results for $\boldtwo-\boldone=(1,1)$ and $(1,-1)$ are 
\begin{align}
(\hat{H}_{\textrm{eff}})_{\boldone \boldtwo}
=
&-\frac{8}{9}
\frac{(t_{12}^{\prime})^{2}}{U-J^{\prime}}
(\frac{1}{4}\hat{n}_{\boldone}\hat{n}_{\boldtwo}
+\hat{\boldJ}_{\boldone}\cdot \hat{\boldJ}_{\boldtwo}
-2 \hat{J}^{z}_{\boldone}\cdot \hat{J}^{z}_{\boldtwo})\notag\\
&-\frac{4}{9}
\frac{(t_{12}^{\prime})^{2}}{U^{\prime}-J_{\textrm{H}}}
(\frac{7}{4}\hat{n}_{\boldone}\hat{n}_{\boldtwo}
+3\hat{J}^{z}_{\boldone}\hat{J}^{z}_{\boldtwo})\notag\\
&-\frac{4}{9}
\frac{(t_{12}^{\prime})^{2}}{U^{\prime}+J_{\textrm{H}}}
(\frac{3}{4}\hat{n}_{\boldone}\hat{n}_{\boldtwo}
-2\hat{\boldJ}_{\boldone}\cdot \hat{\boldJ}_{\boldtwo}
+\hat{J}^{z}_{\boldone}\hat{J}^{z}_{\boldtwo})\notag\\
&-\frac{4}{27}
(\frac{(t_{33}^{\prime})^{2}}{U+2J^{\prime}}+\frac{2(t_{33}^{\prime})^{2}}{U-J^{\prime}})
(\frac{1}{4}\hat{n}_{\boldone}\hat{n}_{\boldtwo}
-\hat{\boldJ}_{\boldone}\cdot \hat{\boldJ}_{\boldtwo})\notag\\
&-\frac{4}{9}
\frac{(t_{33}^{\prime})^{2}}{U^{\prime}-J_{\textrm{H}}}
(\frac{3}{4}\hat{n}_{\boldone}\hat{n}_{\boldtwo}
-\hat{J}^{z}_{\boldone}\hat{J}^{z}_{\boldtwo})\notag\\
&-\frac{4}{9}
\frac{(t_{33}^{\prime})^{2}}{U^{\prime}+J_{\textrm{H}}}
(\frac{1}{4}\hat{n}_{\boldone}\hat{n}_{\boldtwo}
+\hat{J}^{z}_{\boldone}\hat{J}^{z}_{\boldtwo})\notag\\
&-\frac{8}{9}
\frac{t_{12}^{\prime}t_{33}^{\prime}}{U^{\prime}-J_{\textrm{H}}}
(\hat{J}^{x}_{\boldone}\hat{J}^{y}_{\boldtwo}+\hat{J}^{y}_{\boldone}\hat{J}^{x}_{\boldtwo})\notag\\
&+\frac{8}{9}
\frac{t_{12}^{\prime}t_{33}^{\prime}}{U^{\prime}+J_{\textrm{H}}}
(\hat{J}^{x}_{\boldone}\hat{J}^{y}_{\boldtwo}+\hat{J}^{y}_{\boldone}\hat{J}^{x}_{\boldtwo}),
\label{eq:Heff-12-110}
\end{align}
and
\begin{align}
(\hat{H}_{\textrm{eff}})_{\boldone \boldtwo}
=&
-\frac{8}{9}
\frac{(t_{12}^{\prime})^{2}}{U-J^{\prime}}
(\frac{1}{4}\hat{n}_{\boldone}\hat{n}_{\boldtwo}
+\hat{\boldJ}_{\boldone}\cdot \hat{\boldJ}_{\boldtwo}
-2 \hat{J}^{z}_{\boldone}\cdot \hat{J}^{z}_{\boldtwo})\notag\\
&-\frac{4}{9}
\frac{(t_{12}^{\prime})^{2}}{U^{\prime}-J_{\textrm{H}}}
(\frac{7}{4}\hat{n}_{\boldone}\hat{n}_{\boldtwo}
+3\hat{J}^{z}_{\boldone}\hat{J}^{z}_{\boldtwo})\notag\\
&-\frac{4}{9}
\frac{(t_{12}^{\prime})^{2}}{U^{\prime}+J_{\textrm{H}}}
(\frac{3}{4}\hat{n}_{\boldone}\hat{n}_{\boldtwo}
-2\hat{\boldJ}_{\boldone}\cdot \hat{\boldJ}_{\boldtwo}
+\hat{J}^{z}_{\boldone}\hat{J}^{z}_{\boldtwo})\notag\\
&-\frac{4}{27}
(\frac{(t_{33}^{\prime})^{2}}{U+2J^{\prime}}+\frac{2(t_{33}^{\prime})^{2}}{U-J^{\prime}})
(\frac{1}{4}\hat{n}_{\boldone}\hat{n}_{\boldtwo}
-\hat{\boldJ}_{\boldone}\cdot \hat{\boldJ}_{\boldtwo})\notag\\
&-\frac{4}{9}
\frac{(t_{33}^{\prime})^{2}}{U^{\prime}-J_{\textrm{H}}}
(\frac{3}{4}\hat{n}_{\boldone}\hat{n}_{\boldtwo}
-\hat{J}^{z}_{\boldone}\hat{J}^{z}_{\boldtwo})\notag\\
&-\frac{4}{9}
\frac{(t_{33}^{\prime})^{2}}{U^{\prime}+J_{\textrm{H}}}
(\frac{1}{4}\hat{n}_{\boldone}\hat{n}_{\boldtwo}
+\hat{J}^{z}_{\boldone}\hat{J}^{z}_{\boldtwo})\notag\\
&+\frac{8}{9}
\frac{t_{12}^{\prime}t_{33}^{\prime}}{U^{\prime}-J_{\textrm{H}}}
(\hat{J}^{x}_{\boldone}\hat{J}^{y}_{\boldtwo}+\hat{J}^{y}_{\boldone}\hat{J}^{x}_{\boldtwo})\notag\\
&-\frac{8}{9}
\frac{t_{12}^{\prime}t_{33}^{\prime}}{U^{\prime}+J_{\textrm{H}}}
(\hat{J}^{x}_{\boldone}\hat{J}^{y}_{\boldtwo}+\hat{J}^{y}_{\boldone}\hat{J}^{x}_{\boldtwo}),
\label{eq:Heff-12-1m10}
\end{align}
respectively. 
In particular, 
for $J_{\textrm{H}}=0$, $J^{\prime}=0$, and $U^{\prime}=U$, 
$(\hat{H}_{\textrm{eff}})_{\boldone \boldtwo}$ for $\boldtwo-\boldone=(1,\pm 1)$ becomes
\begin{align}
(\hat{H}_{\textrm{eff}})_{\boldone \boldtwo}
=
&-\frac{4}{3}
\frac{(t_{12}^{\prime})^{2}}{U}
\hat{n}_{\boldone}\hat{n}_{\boldtwo}
-\frac{4}{9}
\frac{(t_{33}^{\prime})^{2}}{U}
(\frac{5}{4}\hat{n}_{\boldone}\hat{n}_{\boldtwo}
-\hat{\boldJ}_{\boldone}\cdot\hat{\boldJ}_{\boldtwo}).\label{eq:Heff-12-NNN}
\end{align}

Combining Eqs. (\ref{eq:Heff-12-x-J0}), (\ref{eq:Heff-12-y-J0}), 
and (\ref{eq:Heff-12-NNN}) together 
and neglecting the product terms of the hole-density operators, 
we finally obtain Eq. (\ref{eq:Heff}). 

\section{Derivation of Eq. (\ref{eq:Heff-MFA})}

In this appendix, 
I derive Eq. (\ref{eq:Heff-MFA}). 
The derivation consists of three steps. 

First, 
we rewrite $\hat{H}_{\textrm{eff}}$ in Eq. (\ref{eq:Heff-MFA}) as 
\begin{align}
\hat{H}_{\textrm{eff}}
=\sum\limits_{\boldi,\boldj}
\sum\limits_{\alpha,\beta=x,y,z}
J_{\boldj \boldi}^{\alpha \beta}
\hat{J}_{\boldi}^{\beta}\hat{J}_{\boldj}^{\alpha},\label{eq:Heff-compact-r}
\end{align}
where $\textstyle\sum_{\boldi,\boldj}$ is the summations of $\boldi$ and $\boldj$ 
for all $N$ sites. 
From Eq. (\ref{eq:Heff}), 
we can explicitly write down $J_{\boldj \boldi}^{\alpha \beta}$ 
by recalling 
$t_{ab;\boldi\boldj}^{(\textrm{even})}=t_{ab;\boldj\boldi}^{(\textrm{even})}$, 
$t_{ab;\boldi\boldj}^{(\textrm{odd})}=-t_{ab;\boldj\boldi}^{(\textrm{odd})}$, 
$\textstyle\sum_{\langle \boldi,\boldj \rangle_{x}}\cdots 
=\frac{1}{2}\textstyle\sum_{\boldi,\boldj}\cdots$, 
$\textstyle\sum_{\langle \boldi,\boldj \rangle_{y}}\cdots 
=\frac{1}{2}\textstyle\sum_{\boldi,\boldj}\cdots$, and 
$\textstyle\sum_{\langle\langle \boldi,\boldj \rangle\rangle}\cdots
=\textstyle\sum_{\boldi,\boldj}\cdots$. 
For example, 
$J_{\boldj \boldi}^{\alpha \beta}$ for $\boldj-\boldi=(\pm 1,0)$ and 
$\alpha=\beta=x$ is $\frac{1}{2}(J_{0}-J_{\textrm{odd}})$; 
$J_{\boldj \boldi}^{\alpha \beta}$ for $\boldj-\boldi=(0,\pm 1)$, 
$\alpha=z$, and $\beta=y$ is $\mp\frac{1}{2}D$; 
$J_{\boldj \boldi}^{\alpha \beta}$ for $\boldj-\boldi=(\pm 1,\pm 1)$ and 
$\alpha=\beta=z$ is $J_{2}$. 

Second, 
we apply the mean-field approximation (for example see Ref. \onlinecite{spiral}) 
to Eq. (\ref{eq:Heff-compact-r}) 
by using $\hat{J}_{\boldj}^{\alpha}\approx \langle \hat{J}_{\boldj}^{\alpha}\rangle$, 
and derive the energy at absolute zero of temperature 
in order to determine the ground state. 
As a result, 
the energy is given by 
\begin{align}
\langle \hat{H}_{\textrm{eff}}\rangle
=\sum\limits_{\boldi,\boldj}
\sum\limits_{\alpha,\beta=x,y,z}
J_{\boldj \boldi}^{\alpha \beta}
\langle \hat{J}_{\boldi}^{\beta}\rangle 
\langle \hat{J}_{\boldj}^{\alpha}\rangle .\label{eq:Heff-exp-r}
\end{align} 
Since the mean-field approximation neglects the fluctuations, 
the magnitude of $\langle \hat{J}_{\boldj}^{\alpha}\rangle$ 
should be equal to $J=\frac{1}{2}$ for all $N$ sites. 
Namely, 
$\langle \hat{J}_{\boldj}^{\alpha}\rangle$ in Eq. (\ref{eq:Heff-exp-r}) 
should satisfy the hard-pseudospin constraints, 
\begin{align}
\frac{1}{4}
=&
\frac{1}{N}\sum\limits_{\boldi}
\sum\limits_{\alpha}|\langle \hat{J}_{\boldi}^{\alpha}\rangle|^{2}.\label{eq:constraint-r}
\end{align}

Third, 
by using the Fourier transformation of $\langle \hat{J}_{\boldj}^{\alpha}\rangle$, 
we can rewrite $\langle \hat{H}_{\textrm{eff}}\rangle$ in a quadratic form, 
Eq. (\ref{eq:Heff-MFA}), 
with Eqs. (\ref{eq:Jqxx}){--}(\ref{eq:Jqyz}). 
In addition, 
we can rewrite Eq. (\ref{eq:constraint-r}) as Eq. (\ref{eq:constraint-q}). 
After determining the lowest eigenvalue of Eq. (\ref{eq:Heff-MFA}), 
we should check whether the eigenvector satisfies 
Eq. (\ref{eq:constraint-q}).

\end{document}